\documentclass{aa}
\usepackage[varg]{txfonts}
\usepackage{graphicx}
\usepackage{natbib}
\usepackage{subfig}
\usepackage{hyperref}

\bibpunct{(}{)}{;}{a}{}{,} 

\begin{document}

%
%
    \title{High-resolution HI and CO observations of high-latitude intermediate-velocity clouds}
    \titlerunning{HI and CO observations of intermediate-velocity clouds}


   \author{T.~R{\"o}hser \inst{1} 
	   \and
           J.~Kerp \inst{1}
	   \and 
	   N.~Ben Bekhti \inst{1}
	   \and
	   B.~Winkel \inst{2}
           }

   \institute{Argelander-Institut f{\"u}r Astronomie (AIfA), Universit{\"a}t Bonn, Auf dem H{\"u}gel 71, D-53121 Bonn\\
              \email{troehser@astro.uni-bonn.de}
              \and
              Max-Planck-Institut f{\"u}r Radioastronomie (MPIfR), Auf dem H{\"u}gel 69, D-53121 Bonn
              }

   \date{Received ---; accepted ---}

    \abstract
    {Intermediate-velocity clouds (IVCs) are HI halo clouds that are likely related to a Galactic fountain process. In-falling IVCs are candidates for the {re-}accretion of matter onto the Milky Way.}
    {We study the evolution of IVCs at the disk-halo interface, focussing on the transition from atomic to molecular IVCs. We compare an atomic IVC to a molecular IVC and characterise their structural differences in order to {investigate} how molecular IVCs form high above the Galactic plane.}
    {With high-resolution HI observations of the Westerbork Synthesis Radio Telescope and $^{12}$CO(1$\rightarrow$0) and $^{13}$CO(1$\rightarrow$0) observations with the IRAM 30\,m telescope, we analyse the small-scale structures within the two clouds. By correlating HI and far-infrared (FIR) dust continuum emission from the \textit{Planck} satellite, the distribution of molecular hydrogen (H$_2$) is estimated. We conduct a detailed comparison of the HI, FIR, and CO data and study variations of the $X_\mathrm{CO}$ conversion factor.}
    {The atomic IVC {does not disclose} detectable CO emission. {The atomic small-scale structure, as revealed by the high-resolution HI data, shows low peak HI column densities and low HI fluxes as compared to the molecular IVC}. The molecular IVC exhibits a rich molecular structure and most of the CO emission is observed at the eastern edge of the cloud. There is observational evidence that the molecular IVC is in a transient {and, thus, non-equilibrium} phase. The average $X_\mathrm{CO}$ factor is close to the canonical value of the Milky Way disk.}
    {We propose that the two IVCs represent different states in a gradual transition from atomic to molecular clouds. The molecular IVC {appears to be more} condensed allowing the formation of H$_2$ and CO in shielded regions all over the cloud. Ram pressure {may} accumulate gas and thus facilitate the formation of H$_2$. We {show} evidence that the atomic IVC will evolve also into a molecular IVC in a few Myr.}


   \maketitle

\section{Introduction}
\label{sec:introduction}

\begin{table*}[!t]
  \caption{Characteristics of the different data sets used in this study.
  For IRAM and \textit{Planck}, the data is gridded to FITS-maps with a Gaussian kernel, degrading the spatial resolution slightly. The angular resolution is that of the final gridded maps.}  
  \label{tab:data}
  \small
  \centering
  \begin{tabular}{cccccc}
    \hline\hline
     {Data} & $\nu$ & {Angular resolution} & {Spectral channel width} & {Noise} & Reference \\
     & [GHz] & & [km\,s$^{-1}$] & & \\
     \hline
      EBHIS & 1.42 & 10.8\arcmin & 1.29 & 90\,mK & (1) \\
      WSRT aIVC & 1.42 & $75.1\arcsec\times23.0\arcsec$ & 1.03 & 1.1\,mJy\,(beam)$^{-1}$ & (3) \\
      WSRT mIVC & 1.42 & $49.1\arcsec\times17.9\arcsec$ & 1.03 & 1.4\,mJy\,(beam)$^{-1}$ & (3) \\
      \textit{Planck} $\tau_{353}$ & 353 & 5.27\arcmin & -- & -- & (2) \\      
      IRAM FTS $^{12}$CO(1$\rightarrow$0) & 115.27 & 23.0\arcsec & 0.53 & {0.20}\,K & (3) \\
      IRAM FTS $^{13}$CO(1$\rightarrow$0) & 110.20 & 24.1\arcsec & 0.53 & {0.10}\,K & (3) \\
      IRAM VESPA $^{13}$CO(1$\rightarrow$0) & 110.20 & 24.1\arcsec & 0.13 & {0.15}\,K & (3) \\
     \hline
  \end{tabular}
  \tablebib{
  (1) \citet{Winkel2016,Kerp2011,Winkel2010}; (2) \citet{Planckcollaboration2014XI}; (3) This work.
  }
\end{table*}

In the evolution of a star-forming galaxy like the Milky Way, a cycle of matter is established by the expulsion from the disk and accretion from the halo \citep[e.g.][]{Ferriere2001}. One of the dominant mechanisms is the Galactic fountain process, {that} is {driven} by massive stars and their feedback onto the Galactic {interstellar medium (ISM)} \citep{Shapiro1976,Bregman1980}: Stellar winds and supernovae expel gas and dust into the Galactic halo where a reservoir of metal-enriched material is sustained. The expelled gas cools down and condenses {eventually into atomic clouds}, which are observable by their emission of HI 21\,cm line emission. These clouds {are thought to} fall back and refuel the Milky Way disk \citep[e.g.][]{Putman2012}.

Usually, HI halo clouds are identified by their observed radial velocities which are incompatible with simple models of Galactic rotation \citep[e.g.][]{Wakker1991}. \citet{Wakker2001} uses a velocity range relative to the {local standard of rest} (LSR) between $40\,$km\,s$^{-1}\lesssim |v_{\mathrm{LSR}}| \lesssim90\,$km\,s$^{-1}$ to define intermediate-velocity clouds (IVCs). Most of the IVCs show metallicities close to solar, contain dust as seen by their far-infrared (FIR) continuum emission, and have distances below 5\,kpc \citep[e.g.][]{Wakker2001}. All these properties favour a connection of IVCs to Galactic fountains \citep{Bregman2004,Putman2012,Sancisi2008}.

For the evolution of IVCs in the Galactic halo, not only their atomic but also their molecular content is important. The most efficient formation mechanism of molecular hydrogen (H$_2$) is the formation on the surfaces of dust grains \citep{Hollenbach1971b}. Dust is present in IVCs as is evident from their FIR emission \citep[e.g.][]{Planckcollaboration2011XXIV}. As a likely product of a Galactic fountain, not only gas but also dust is expelled into the Galactic halo \citep{Putman2012}. Molecular hydrogen is observed in IVCs, either as a diffuse low column density component with $N_{\mathrm{H}_{2}} = 10^{14}-10^{16}\,\mathrm{cm}^{-2}$ \citep{Richter2003,Wakker2006} or as intermediate-velocity molecular clouds (IVMCs). {These clouds contain} significant molecular fractions such that $^{12}$CO(1$\rightarrow$0) emission is detectable \citep{Magnani2010}.

{Which state the clouds are in when they impact the disk at the end of the fountain cycle is important}. If the clouds are destroyed and ionised, they cannot contribute to star formation for which cold gas is required \citep{Putman2012}. In-falling cold and dense clouds may integrate into the disk and feed star formation or even trigger the formation of molecular clouds and stars, for instance in the Gould Belt \citep{Comeron1992}.

\citet{Roehser2014} propose a natural evolutionary sequence from pure atomic to molecular IVCs in the fountain cycle. During the in-fall of IVCs, ram pressure perturbs the clouds by their motion through the surrounding halo medium. Enhanced pressure leads to the faster formation of H$_2$ which is related to the compression and accumulation of the gas \citep{Guillard2009,Hartmann2001}. These effects are most important during the final stages of accretion of the clouds because the surrounding halo medium is densest.

\citet{Roehser2014} base their discussion on two prototypical IVCs at high Galactic latitudes that show an in-falling motion. These IVCs appear as twins in HI single-dish data but are completely different in terms 
of the correlation with the FIR dust emission: One cloud is purely atomic, the other is a rare IVMC. {\citet{Roehser2014} focus on the HI-FIR correlation on large angular scales.}

Here, we present new high-resolution observations of these two clouds in HI and CO. The different chemical properties of the two clouds are expected to be imprinted in their spatial small-scale structure \citep{Hennebelle2012}. We study the connection between the atomic and molecular gas. By correlating the HI emission to the FIR dust continuum, we estimate the distribution of H$_2$ {at smaller angular scales as compared to \citet{Roehser2014}}. Thus, variations of the conversion factor between CO and H$_2$, the $X_{\mathrm{CO}}$-factor \citep{Bolatto2013}, are derived for the IVMC.

This paper is organised as follows. In Section \ref{sec:data} we present the data sets that are used in this study. In Section \ref{sec:methods} we describe how we infer H$_2$ column densities. In Section \ref{sec:results} we present the characteristics of the two IVCs that are obtained from the new high-resolution data. In Section \ref{sec:discussion} we discuss our results and {we} summarise in Section \ref{sec:summary}.

\section{Data}
\label{sec:data}

The {basic parameters of the} data sets are compiled in Table \ref{tab:data}. Henceforward, we refer to the atomic IVC as "aIVC" and to the molecular IVC as "mIVC". These are the clouds IVC\,1 and IVC\,2 from \citet{Roehser2014}.

\subsection{HI data}
\label{sec:hi-data}

For our HI analysis we use the new Effelsberg-Bonn HI Survey \citep[EBHIS,][]{Winkel2016,Kerp2011,Winkel2010}, which is complemented by new high-resolution radio interferometric observations {conducted} with the Westerbork-Synthesis Radio Telescope (WSRT). The mIVC is observed with a single {WSRT} pointing, the aIVC with two {WSRT} pointings because of its larger angular extent. Each field of interest was observed for 12 hours to achieve both a good {$uv$-coverage} and sensitivity. The observations were conducted in June 2013.

The WSRT data is calibrated with the software package MIRIAD \citep{Sault1995} and imaged with CASA \citep{McMullin2007}. Only a small fraction of the data {is rejected from the analysis (flagged)} due to contamination with radio-frequency interference. Self-calibration is performed towards the brightest unresolved continuum sources in order to improve the phase solutions. The continuum is subtracted for each pointing separately by fitting a low-order polynomial to the line-free channels. The data is iteratively CLEANed using multiscale-CLEAN in CASA {using the HI data cubes from EBHIS as input models.} Masks are set around the emission in order to avoid CLEANing of the strong imaging artefacts. The CLEANing is a crucial step because of the extended HI emission of both IVCs which covers the primary beam of the interferometer {completely}. We apply Briggs weighting with a robust parameter of two which is close to natural weighting in order to achieve the best sensitivity \citep{Briggs1999}. For the aIVC, the two pointings are imaged separately and combined into a mosaic image. The interferometric images are corrected for the {sensitivity pattern of the} primary beam {amplifying the noise at the field boundaries. Thus, we blank the interferometric data below a primary-beam sensitivity level of 0.25. We choose this cut-off value because most HI emission is preserved.}

The calibrated and imaged interferometric data is supplemented by the EBHIS data to fill in the missing spacings. This is of great importance for nearby objects with extended HI emission that is not well recovered by the interferometer \citep[e.g.][]{Stanimirovic2002}. For the combination we use the method of \citet{Faridani2014}, which combines the science-ready data in the image domain. {We tested that their approach gives similar results as the {feathering task} implemented in CASA.} The properties of the final HI data cubes are {summarised} in Table \ref{tab:data}.

{However, for our targets the combination is affected by systematic biases because there is lots of HI emission outside the mapped WSRT pointings.} This introduces uncertainties in the analysis of the two clouds because the small-scale structures of the surrounding HI gas is unknown. {Nevertheless the combined EBHIS and WSRT HI data is used for the HI-$\tau$ correlation (Sects.~\ref{sec:atomic-ivc-hi} and \ref{sec:molecular-ivc-hi}) in order to advance in angular resolution as compared to the study of \citet{Roehser2014}. For the analysis of the structural properties of the clouds (Sect.~\ref{sec:comparison}) the EBHIS and WSRT data are evaluated separately.}

\subsection{CO data}
\label{sec:co-data}

With the IRAM 30\,m telescope at the Pico Veleta (Spain) we mapped both IVCs in $^{12}$CO(1$\rightarrow$0) and $^{13}$CO(1$\rightarrow$0) emission applying on-the-fly (OTF) mode with position switching. We used the single-pixel receiver EMIR \citep{Carter2012} in combination with the FTS spectrometer \citep{Klein2012} in wide mode offering 8\,GHz of total bandwidth at a spectral resolution of 195\,kHz. The spectral band was chosen such that both the $^{13}$CO(1$\rightarrow$0) transition at 110.20\,GHz and the $^{12}$CO(1$\rightarrow$0) transition at 115.27\,GHz were observed simultaneously. In addition, the backend VESPA\footnote{http://www.iram.fr/IRAMFR/TA/backend/veleta/vespa/index.htm} was used for high spectral resolution spectroscopy {providing} 50\,kHz centred on the $^{13}$CO(1$\rightarrow$0) line. 

The OTF stripes were positioned such that the data fulfils Nyquist sampling. The scanning direction was either along right ascension {(R.A.)} or declination {(Dec.)} with an integration time of {1\,s}. Regularly, every 1--2\,min an {emission-free} reference position near the scanned field is observed. About every 10\,min the data is calibrated by observing the sky emission at the reference position and the counts from loads at ambient and cold temperatures. The average opacities were $\tau\simeq0.2-0.3$ at 115\,GHz. In total, 89\,h of observing time was scheduled May 31 2014 -- June 3 2014 and August 28 -- September 04 2014. In the first session lots of observation time was lost due to bad weather {conditions}.

The mean elevation of the two clouds was about 60$^\circ$. For the mIVC we covered those regions which contain most of the FIR emission. Jorge L. Pineda kindly provided us with unpublished IRAM 30\,m observations of the central core of the cloud. In total, the observations consist of 223481 individual spectra. For the aIVC a field centred on the largest HI column densities was observed, in total 24946 spectra.

The data reduction was performed with the Grenoble Image and Line Data Analysis Software (GILDAS)\footnote{http://www.iram.fr/IRAMFR/GILDAS}. The OTF data was re-calibrated with the Multichannel Imaging and Calibration Software for Receiver Arrays (MIRA)\footnote{http://www.iram.fr/IRAMFR/GILDAS/doc/html/mira-html/mira.html} in order to improve on the standard automatic calibration scheme. With the Continuum and Line Analysis Single-dish Software (CLASS)\footnote{http://www.iram.fr/IRAMFR/GILDAS/doc/html/class-html/class.html}, we extract 300 channels centred on the $^{12}$CO(1$\rightarrow$0) and $^{13}$CO(1$\rightarrow$0) and correct for the beam efficiency of the telescope.

With self-written python\footnote{http://www.python.org/}-scripts, we fit a baseline of fifth order to the line-free channels of each spectrum and subtract the baseline polynomial from the entire spectrum. These spectra are gridded to FITS\footnote{http://fits.gsfc.nasa.gov/} data cubes with pixel size of ${8}\arcsec$ using a Gaussian kernel of 1/3 of the telescope beam. Prior gridding, the additional data from Jorge L. Pineda data is spectrally smoothed and re-gridded in order to match the FTS data. The properties of the final CO data cubes are given in Table \ref{tab:data}.

\subsection{FIR data}
\label{sec:fir-data}

In the FIR we use the all-sky dust model from \citet{Planckcollaboration2014XI} who fit modified black-bodies to the dust spectrum at 353\,GHz, 545\,GHz, 857\,GHz, and 3000\,GHz. The amplitude of the emission strength is given by the dust optical depth at 353\,GHz $\tau_{353}$ to which we refer to in the following as $\tau$. The dust model is calculated {for an angular} resolution of $5\arcmin$, which is provided in HEALPix\footnote{http://healpix.sourceforge.net/} format \citep{Gorski2005}. For visualisation, we grid the HEALPix data to FITS-maps by applying a Gaussian kernel which degrades the spatial resolution slightly (Table \ref{tab:data}).

\section{Methods}
\label{sec:methods}

There is a linear correlation between the total hydrogen column density $N_\mathrm{H}$ and the FIR dust continuum emission $I_\nu$ at frequency $\nu$ \citep[e.g.][]{Boulanger1996}. Equivalently, the dust emission strength can be expressed as the amplitude $\tau$ of a modified black-body \citep[e.g.][]{Planckcollaboration2014XI}. We use $\tau$ as the measure for the dust emission because it combines information from four different frequency bands reducing variations of the emissivity in individual bands.

We write the correlation as
\begin{equation}
 \label{eq:h-ir-corr}
 \tau = R + \epsilon \times N_\mathrm{H} \simeq R + \epsilon \times (N_\mathrm{HI} + 2N_{\mathrm{H}_2})
\end{equation}
with a constant offset $R$ and the dust emissivity $\epsilon$. The ionised hydrogen column density is neglected here because of its low {contribution} \citep[e.g.][]{Lagache2000}. 

In order to model the observed FIR emission best, we adopt a two-component model consisting of a foreground and the IVC located beyond. These two components may have dust-to-gas ratios that are generally different. For the two IVCs of interest the foreground HI column density is low and {its} molecular fraction negligible. Thus, we write
\begin{equation}
 \label{eq:h-ir-corr2}
  \tau \simeq R + \epsilon^{\mathrm{local}} \times N_\mathrm{HI}^{\mathrm{local}} + \epsilon^{\mathrm{IVC}} \times (N_\mathrm{HI}^{\mathrm{IVC}} + 2N_{\mathrm{H}_2}^{\mathrm{IVC}}).
\end{equation}
By the inversion of this relation one obtains an expression for the non-linearly correlating gas component, which is {attributed to} molecular hydrogen, assuming a constant dust-to-gas ratio in both the local foreground and the IVC. Rearranging to the H$_2$ column density of the IVC:
\begin{equation}
 \label{eq:nh2-ivc}
  N_{\mathrm{H}_{2}}^{\mathrm{IVC}} \simeq \frac{1}{2} \left( \frac{\tau-(R+\epsilon^{\mathrm{local}}\times N_{\mathrm{HI}}^{\mathrm{local}})}{\epsilon^{\mathrm{IVC}}}-N_{\mathrm{HI}}^{\mathrm{IVC}} \right).
\end{equation}
The linear parameters $R$, $\epsilon^{\mathrm{local}}$, and $\epsilon^{\mathrm{IVC}}$ are estimated from fits to the linear part of the HI-$\tau$ correlation in which the influence of H$_2$ is negligible also {for} the IVC {gas}. For increasing HI column density more {and more} atomic hydrogen turns molecular, which leads to the FIR excess \citep[e.g.][]{Desert1988}. The important limitation is the determination of the linear parameters in Eq.~\eqref{eq:h-ir-corr2}.

Instead of fitting the two-component linear function (Eq.~\ref{eq:h-ir-corr2}) up to a fixed HI column density threshold \citep[e.g.][]{Reach1998}, we iteratively remove data points from the HI-$\tau$ correlation that deviate too much from an initial fit. The deviation is calculated as the standard deviation of the residual emission, similar to \citet{Planckcollaboration2011XXIV}. We start with the entire correlation data and subsequently exclude all points that deviate by more than $5\sigma$ from this initial estimate. For the new set of data points, Eq.~\ref{eq:h-ir-corr2} is fitted again, the deviation evaluated and data points excluded. This is continued in a loop until each remaining data point is consistent with the 5$\sigma$ threshold. We repeat the loop with a $4\sigma$, $3\sigma$, and $2\sigma$ rejection criterion. Each loop is continued until all data points are consistent with the respective deviation criterion.

\renewcommand{\arraystretch}{1.5}	
\begin{table*}[!t]
  \caption{Fitted linear parameters for the aIVC and mIVC. The values are derived by fitting the two-component model (Eq.~\ref{eq:h-ir-corr2}) to the HI-$\tau$ correlation plots {of the combined low- and high-resolution HI and FIR data} (Fig.~\ref{fig:hi-ir-corr}) using the iterative fitting method as described in Section \ref{sec:methods}.}  
  \label{tab:hi-ir-coeff}
  \small
  \centering
  \begin{tabular}{ccc|ccc}
    &  aIVC & & & mIVC & \\
    \hline\hline
    $R$ $[10^{-6}]$ & $\epsilon^{\mathrm{local}}$ $[10^{-26}\,\mathrm{cm}^{2}]$ & $\epsilon^{\mathrm{IVC}}$ $[10^{-26}\,\mathrm{cm}^{2}]$ & $R$ $[10^{-6}]$ & $\epsilon^{\mathrm{local}}$ $[10^{-26}\,\mathrm{cm}^{2}]$ & $\epsilon^{\mathrm{IVC}}$ $[10^{-26}\,\mathrm{cm}^{2}]$ \\  
    \hline
    $0.12\pm0.01$ & $0.56\pm0.01$ & $0.391\pm0.003$ & $-0.96\pm0.02$ & $1.47\pm0.03$ & $1.17\pm0.01$\\ 
    \hline
  \end{tabular}
\end{table*}
\renewcommand{\arraystretch}{1}

{We investigate a test case of a FIR excess cloud with higher dust emissivity at large HI column densities in addition to some linearly-correlating foreground, similar to the molecular cloud studied by \citep{Lenz2015}. The estimated linear parameters from the iterative method are compared to those with a fixed upper HI column density threshold. The iterative removal performs significantly better (compare also with Fig.~\ref{fig:hi-ir-corr}). Although, the estimated slope is still slightly in excess of the true one because of the smooth transition between the linearly correlating foreground and the steeper-correlating cloud. In addition, the best cut-off column density is not well constrained.}

For the HI-$\tau$ correlation we have different {HI} data sets with high and low spatial resolution (Table \ref{tab:data}). Thus, one has to be careful to compare the data at the same spatial resolution. We smooth the WSRT data to the angular resolution of \textit{Planck}. The smoothed interferometric data is combined with EBHIS. {Outside the field-of-view of the WSRT the FIR data is smoothed to the angular resolution of EBHIS.}

\section{Results}
\label{sec:results}

\begin{figure*}[!t]
  \centering
  \vspace{-0.5cm}
  \subfloat{\includegraphics[width=0.45\textwidth]{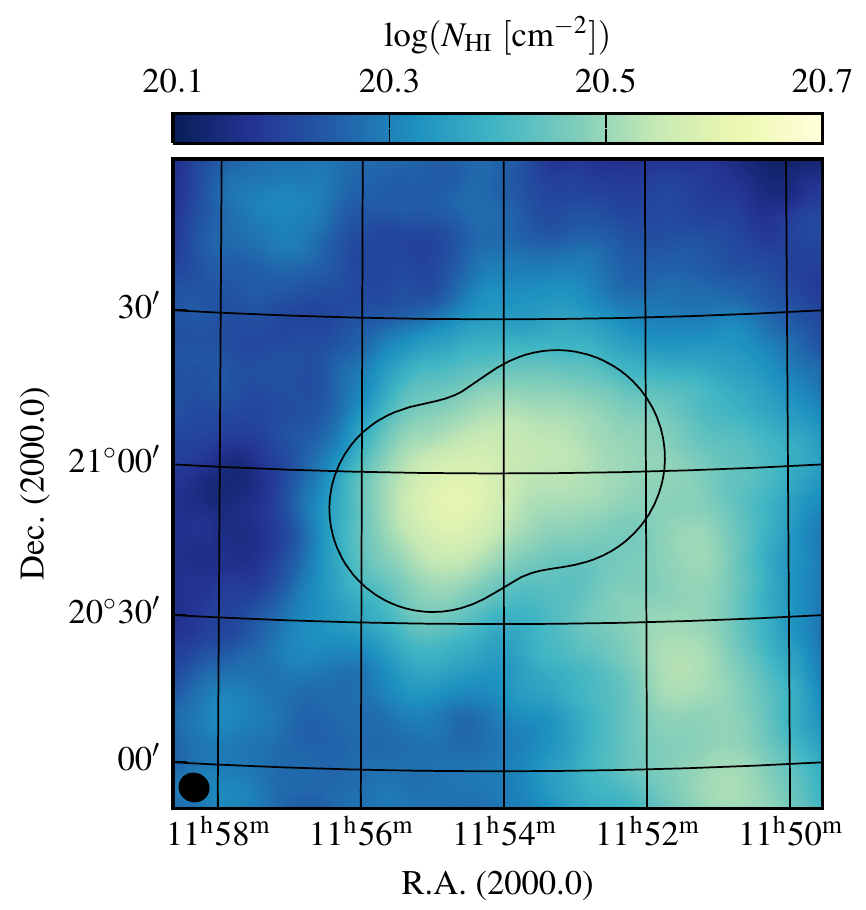}}
  \hspace{0.5cm}
  \subfloat{\includegraphics[width=0.45\textwidth]{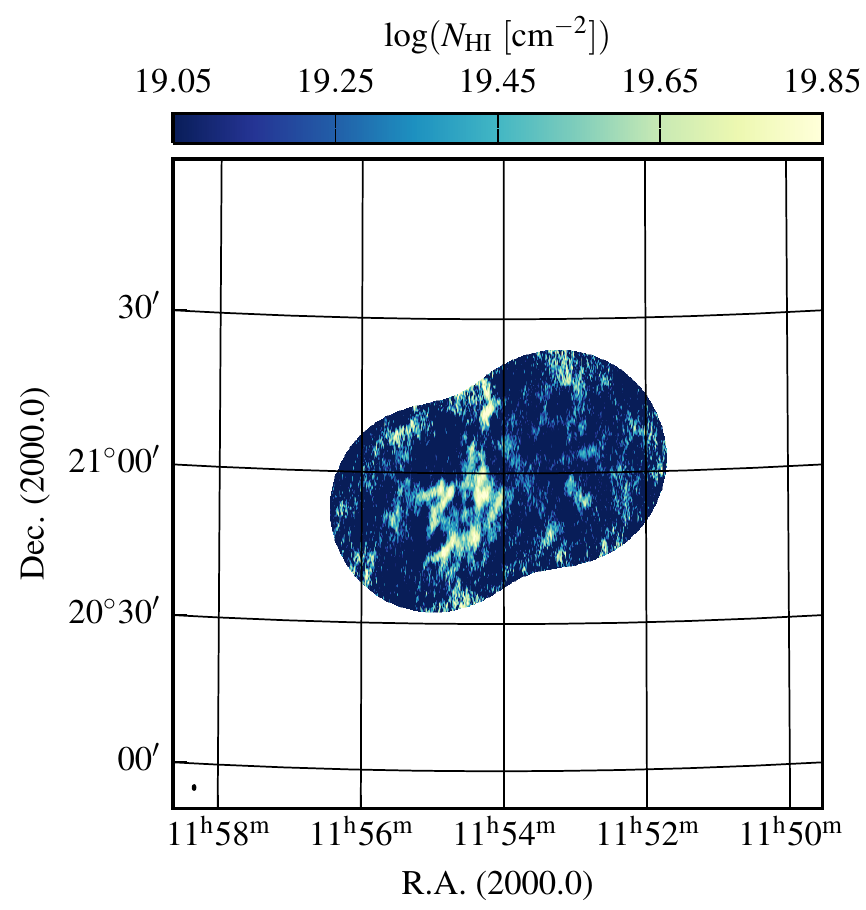}}\\
  \vspace{-0.5cm}
  \subfloat{\includegraphics[width=0.45\textwidth]{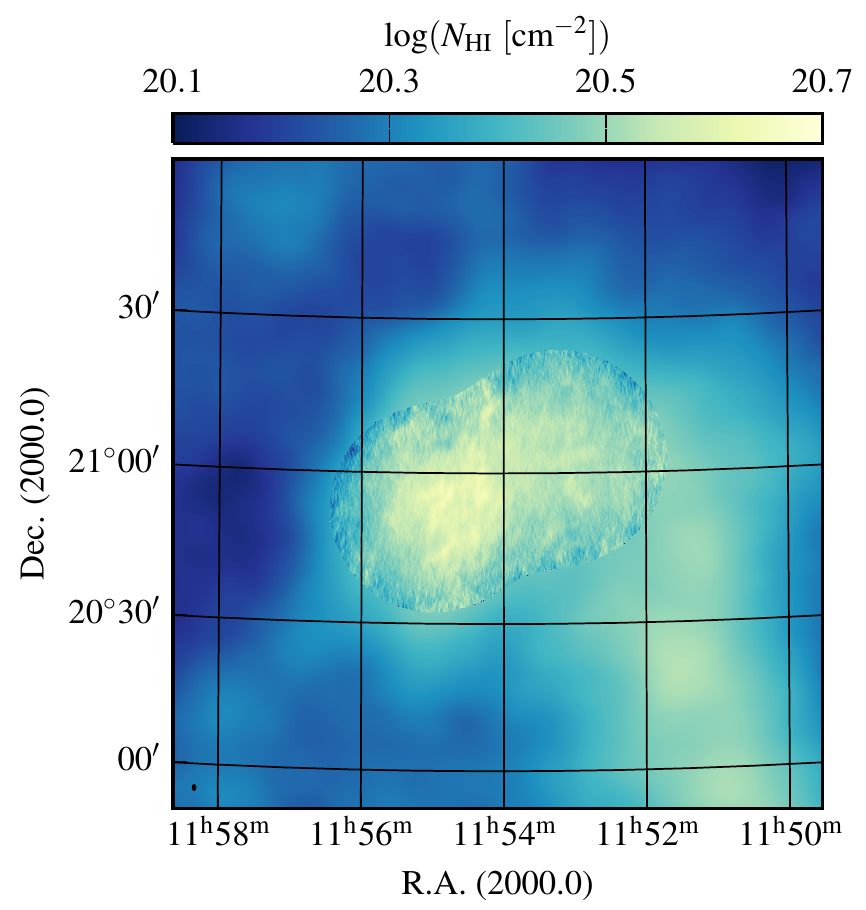}}
  \hspace{0.5cm}
  \subfloat{\includegraphics[width=0.45\textwidth]{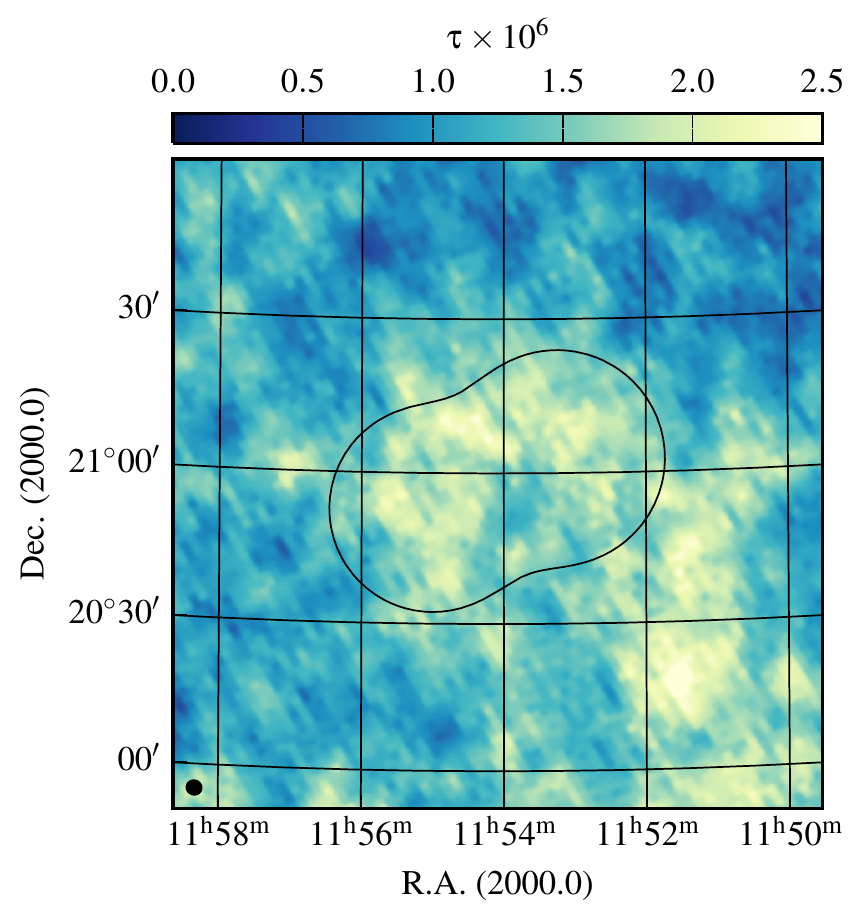}}\\
  \caption{HI column density maps of the aIVC for EBHIS (\textit{top left}), WSRT  (\textit{top right}), and the combined HI map (\textit{bottom left}). The column densities are calculated by integrating over $-77.1\,\mathrm{km}\,\mathrm{s}^{-1}\lesssim v_\mathrm{LSR} \lesssim -20.4\,\mathrm{km}\,\mathrm{s}^{-1}$. The colour-scaling is the same for the EBHIS and the combined map. At the bottom right the \textit{Planck} $\tau$ data is shown. {The black shape marks the position and extent of the two WSRT pointings.} The beam size of each map is given in the bottom-left corner.
  }
  \label{fig:aivc-maps}
\end{figure*}

\subsection{Atomic IVC}
\label{sec:atomic-ivc}

The atomic IVC (aIVC) is located at (R.A., Dec.) $=$ ($11^{\mathrm{h}} 52^{\mathrm{m}}$, $20^{\circ}30\arcmin$) with a radial velocity of $v_\mathrm{LSR}\simeq-35\,\mathrm{km}\,\mathrm{s}^{-1}$. By position and velocity it can be associated with the intermediate-velocity (IV) Spur \citep{Wakker2004}. The cloud has low FIR emission and is thought to be purely atomic on the basis of EBHIS and \textit{Planck} {data} \citep{Roehser2014}.

\subsubsection{HI and FIR observations}
\label{sec:atomic-ivc-hi}

The EBHIS, WSRT, combined HI column density maps, and the \textit{Planck} $\tau$ data are shown in Fig. \ref{fig:aivc-maps}. The column densities are calculated by summing over $-77.1\,\mathrm{km}\,\mathrm{s}^{-1}\lesssim v_\mathrm{LSR} \lesssim -20.4\,\mathrm{km}\,\mathrm{s}^{-1}$.

Both WSRT pointings disclose structures that are unresolved by the single dish. The column density contrast between the EBHIS and WSRT data is large: The peak HI column density in the Effelsberg data is $\sim$$4.0\times10^{20}\,\mathrm{cm}^{-2}$, in the primary-beam corrected WSRT data {only} $\sim$$1.0\times10^{20}\,\mathrm{cm}^{-2}$. Similarly, only about {$\sim$2\%} of the total flux is recovered by the interferometer.
Hence, {only a small fraction of the total HI emission is} associated with the HI small-scale structure of the aIVC. In the {WSRT} HI data cube the brightness temperatures range up to $T_\mathrm{B}\simeq{20}\,$K with line widths as narrow as $\mathrm{FWHM}\simeq{3}\,\mathrm{km}\,\mathrm{s}^{-1}$.

\begin{figure*}[!t]
  \centering
  \includegraphics[width=0.45\textwidth]{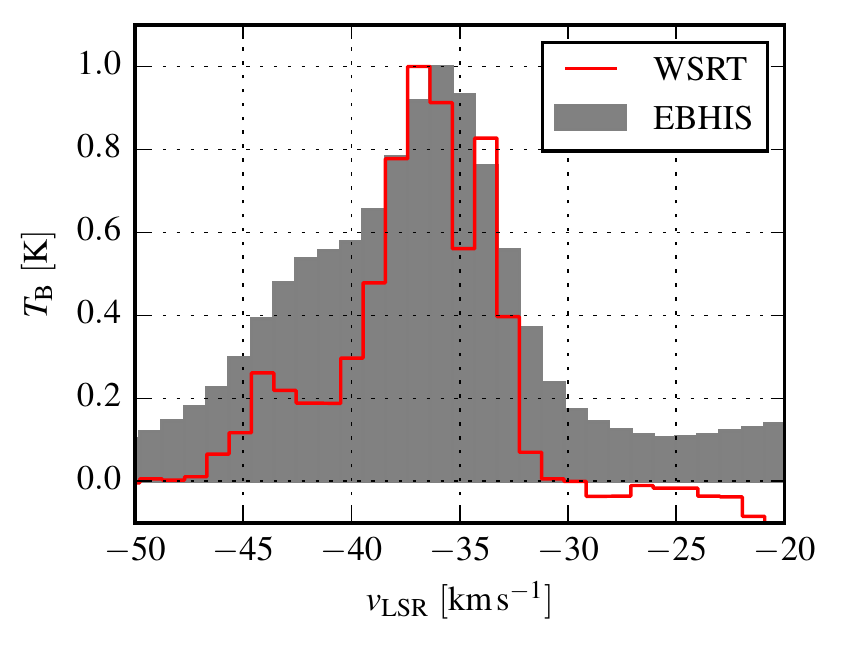}
  \includegraphics[width=0.45\textwidth]{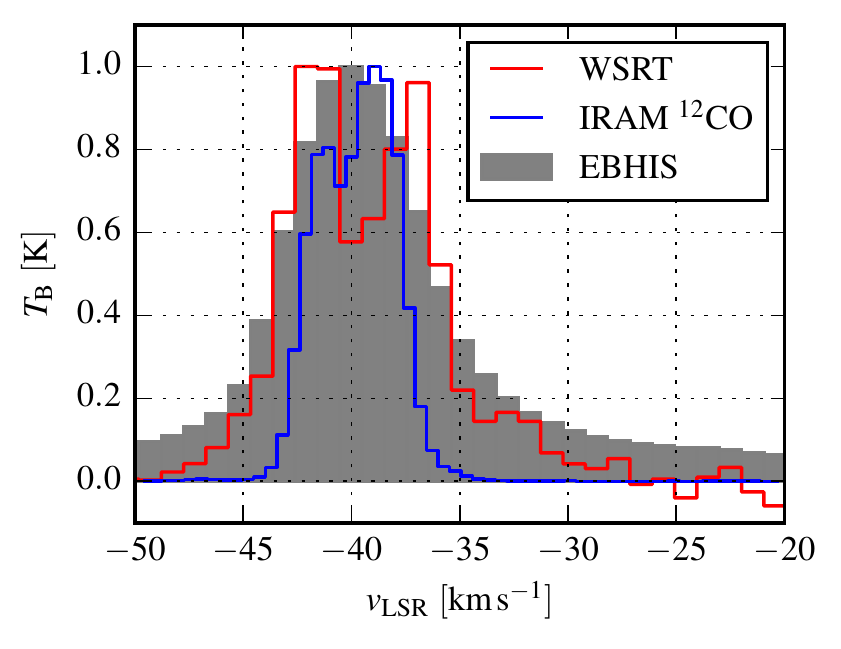}\\
  \caption{{Average HI and $^{12}$CO(1$\rightarrow$0) spectra of the aIVC (\textit{left}) and mIVC (\textit{right}). The average EBHIS spectrum is shown in grey, on top of which the average spectra from the WSRT and IRAM (for the mIVC only) are plotted in red and blue. All spectra are normalised to unity for better comparison.}
  }
  \label{fig:average-spectra}
\end{figure*}

{In Fig.~\ref{fig:average-spectra} (left) we show the average HI spectra for the aIVC. The average EBHIS HI spectrum is plotted in grey and the average WSRT HI in red. While there is only a single broad peak in the EBHIS data, the WSRT resolves a two-peak HI spectrum with a separation of $\sim$$3\,\mathrm{km}\,\mathrm{s}^{-1}$. Spectrally, a third component is apparent near $v_\mathrm{LSR}\simeq-44\,\mathrm{km}\,\mathrm{s}^{-1}$.}

\begin{figure*}[!t]
  \centering
  \subfloat{\includegraphics[width=0.45\textwidth]{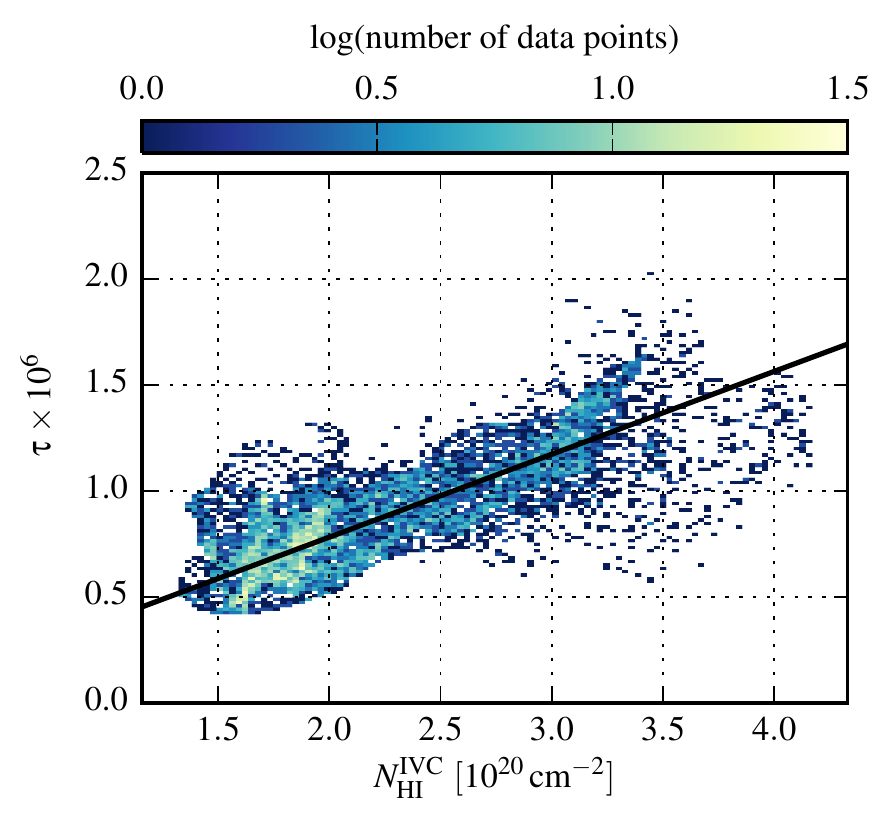}}
  \subfloat{\includegraphics[width=0.45\textwidth]{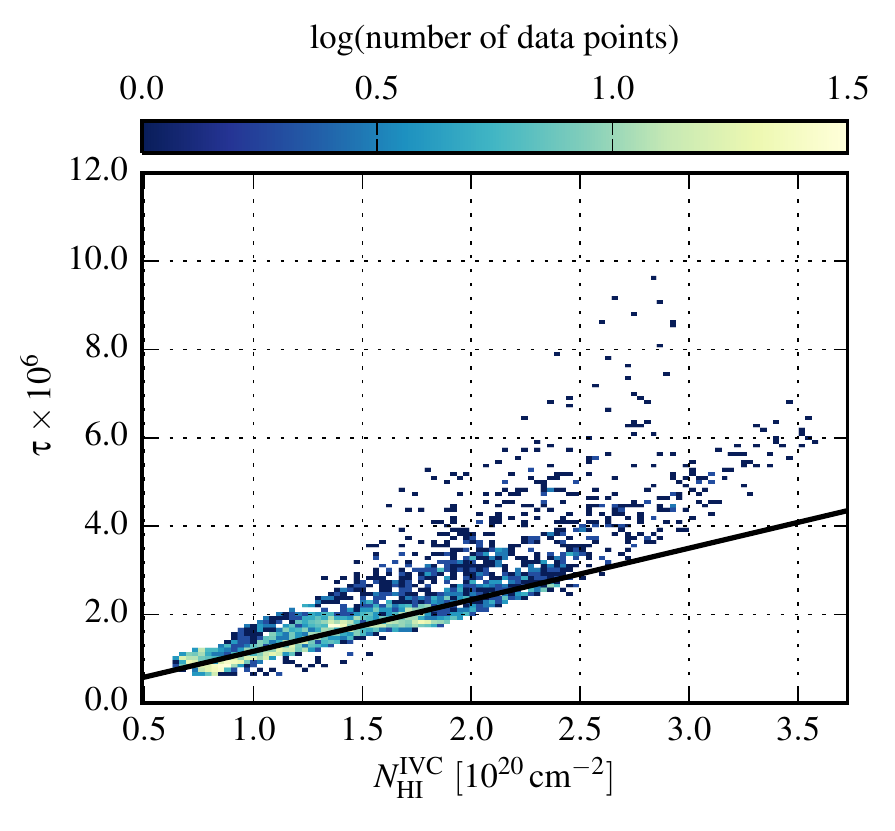}}\\
  \caption{HI-$\tau$ correlation plots for the aIVC (\textit{left}) and mIVC (\textit{right}). The black lines mark the IVC component of the two-component linear model (Eq.~\ref{eq:h-ir-corr2}) using the iterative fitting method (Section \ref{sec:methods}).
  }
  \label{fig:hi-ir-corr}
\end{figure*}

In the FIR the aIVC {is only slightly brighter than the surroundings rendering the cloud difficult to identify by eye}. However, the HI-$\tau$ correlation plot (Fig.~\ref{fig:hi-ir-corr}, left) indicates that the {gas and dust are related to each other}. In the correlation plot {between the combined low- and high-resolution HI and FIR data} the black line marks the IVC contribution to the linear two-component function (Eq.~\ref{eq:h-ir-corr2}). The estimated linear parameters are listed in Table \ref{tab:hi-ir-coeff}.

\subsubsection{IRAM 30\,m CO observations}
\label{sec:atomic-ivc-co}

At the location of the largest HI column densities the aIVC is not detected in $^{12}$CO(1$\rightarrow$0) emission, even when the data is strongly spatially and spectrally smoothed. The 3$\sigma$ detection limit in the smoothed data cube is $\sim$0.02\,K. Using the standard $X_\mathrm{CO}=2\times10^{20}\,\mathrm{cm}^{-2}\,(\mathrm{K}\,\mathrm{km}\,\mathrm{s}^{-1})^{-1}$ \citep{Bolatto2013}, this corresponds to an upper limit of $N_{\mathrm{H}_2}^\mathrm{IVC}\simeq4.3\times10^{18}\,\mathrm{cm}^{-2}$. Hence, the aIVC {can be considered as a purely atomic cloud} with a {negligible} fraction of H$_2$.

\subsection{Molecular IVC}
\label{sec:molecular-ivc}

The molecular IVC (mIVC) is located at (R.A., Dec.) $=$ ($10^{\mathrm{h}}52^{\mathrm{m}}$, $25^{\circ}$) with a radial velocity of $v_\mathrm{LSR}\simeq-40\,\mathrm{km}\,\mathrm{s}^{-1}$. By position and velocity, it is part of the IV Arch \citep{Wakker2004}. As seen by EBHIS and \textit{Planck}, the cloud shows bright FIR emission and a strong FIR excess which is indicative of H$_2$ \citep{Roehser2014}. Pointed CO observations of \citet{Desert1990a} detect the cloud and thus confirm its nature as a molecular cloud.

\subsubsection{HI and FIR observations}
\label{sec:molecular-ivc-hi}

\begin{figure*}[!t]
  \centering
  \vspace{-0.5cm}
  \subfloat{\includegraphics[width=0.45\textwidth]{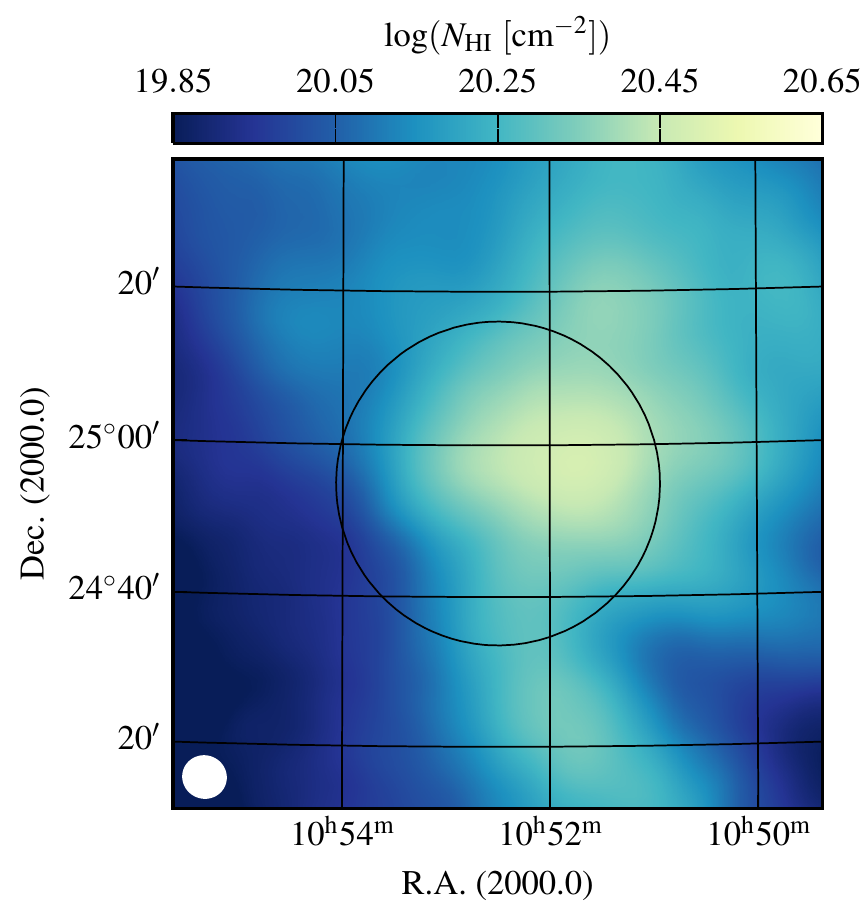}}
  \hspace{0.5cm}
  \subfloat{\includegraphics[width=0.45\textwidth]{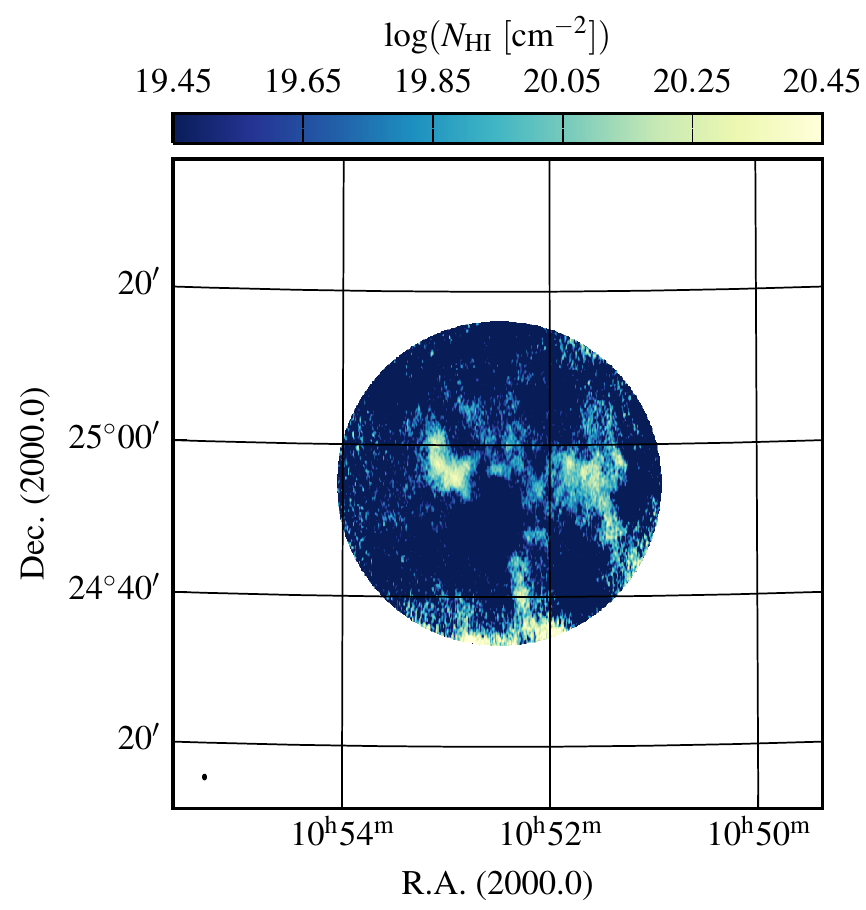}}\\
  \vspace{-0.5cm}
  \subfloat{\includegraphics[width=0.45\textwidth]{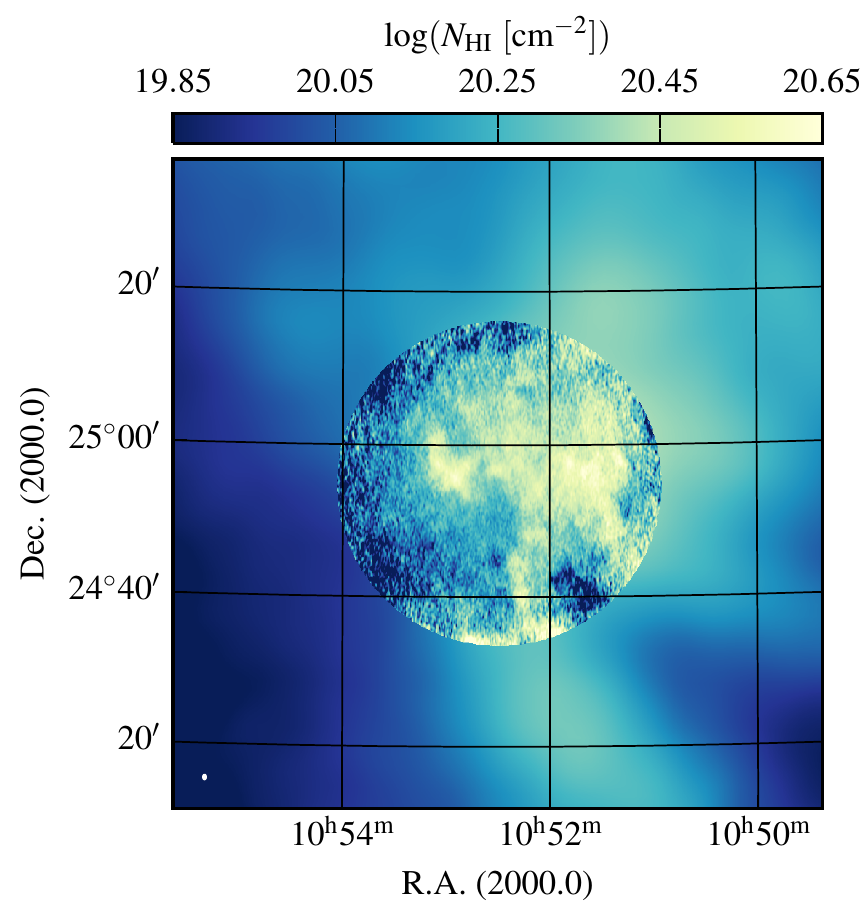}}
  \hspace{0.5cm}  
  \subfloat{\includegraphics[width=0.45\textwidth]{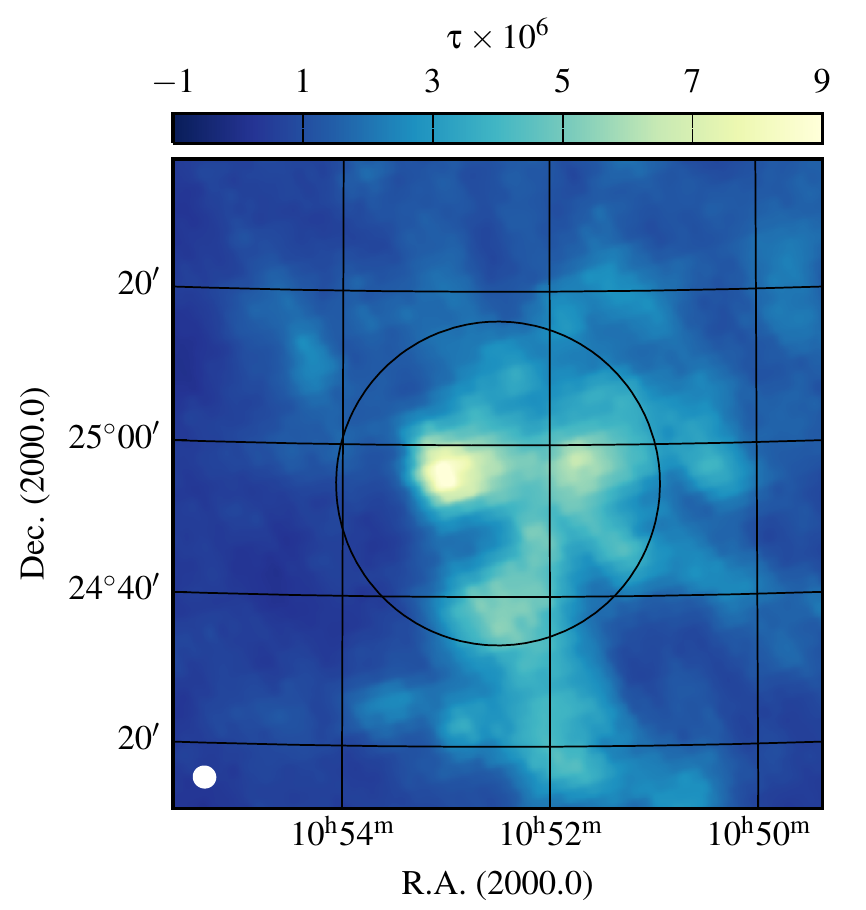}}\\
  \caption{HI column density maps of the mIVC for EBHIS (\textit{top left}), WSRT (\textit{top right}), and the combined HI map (\textit{bottom left}). The column densities are calculated for channels with $-56.5\,\mathrm{km}\,\mathrm{s}^{-1}\lesssim v_\mathrm{LSR} \lesssim -20.4\,\mathrm{km}\,\mathrm{s}^{-1}$. At the bottom right the \textit{Planck} $\tau$ data is shown. {The black circle marks the position and extent of the WSRT pointing.} The colour-scaling is the same for the EBHIS and the combined map. The beam size of each map is given in the bottom-left corner.
  }
  \label{fig:mivc-maps}
\end{figure*}

The observed maps (EBHIS, WSRT, combined HI column density, \textit{Planck} $\tau$) are displayed in Fig.~\ref{fig:mivc-maps}. The column densities are calculated for channels with $-56.5\,\mathrm{km}\,\mathrm{s}^{-1}\lesssim v_\mathrm{LSR} \lesssim -20.4\,\mathrm{km}\,\mathrm{s}^{-1}$.

The peak HI column densities of EBHIS and WSRT are comparable with $\sim$$3.1\times10^{20}\,\mathrm{cm}^{-2}$ and $\sim$$2.4\times10^{20}\,\mathrm{cm}^{-2}$. About {$\sim$5\%} of the total flux is recovered by the interferometer.
In the {WSRT} HI data cube, the brightness temperatures range up to $T_\mathrm{B}\simeq{30}\,$K with line widths as narrow as $\mathrm{FWHM}\simeq3\,\mathrm{km}\,\mathrm{s}^{-1}$.

{In Fig.~\ref{fig:average-spectra} (right) we show the average HI and $^{12}$CO(1$\rightarrow$0) spectra for the mIVC. The average $^{12}$CO(1$\rightarrow$0) spectrum is calculated from the $5\sigma$-masked $^{12}$CO data cubes. The average EBHIS HI spectrum is plotted in grey, the average WSRT HI in red, and the average IRAM $^{12}$CO(1$\rightarrow$0) spectrum in blue. While there is only a single peak in the EBHIS data, the WSRT resolves a two-peak HI spectrum with a peak separation of $\sim$$5\,\mathrm{km}\,\mathrm{s}^{-1}$. In velocity, the CO emission is found close to the HI peak at lower absolute velocity with a slight shift towards larger negative velocities, extending in between the two HI peaks.}

{In the WSRT HI data the largest HI column densities are found} at the eastern edge of the mIVC. {The EBHIS data suggests that most of the HI} is located at the cloud centre towards the west.

The mIVC is prominent in the FIR with a double-peak structure {(Fig.~\ref{fig:mivc-maps}, bottom-right)} that has been noted by \citet{Desert1990a}. {The bright eastern FIR maximum suggests the presence of additional gas that is not atomic.}

The HI-$\tau$ correlation plot for the IVC component (Fig.~\ref{fig:hi-ir-corr}, right) reveals strong FIR excess emission above IVC HI column densities of $\sim$${1.5}\times10^{20}\,\mathrm{cm}^{-2}$. In Fig.~\ref{fig:hi-ir-corr} the black line marks the modelled $\tau$ of the IVC from the fitted linear two-component function (Eq.~\ref{eq:h-ir-corr2}) {using the combined low- and high-resolution HI and FIR data}. The corresponding linear parameters are listed in Table \ref{tab:hi-ir-coeff}. 

\subsubsection{IRAM 30\,m CO observations}
\label{sec:molecular-ivc-co}

\begin{figure*}[!t]
  \centering
  \subfloat{\includegraphics[width=0.5\textwidth]{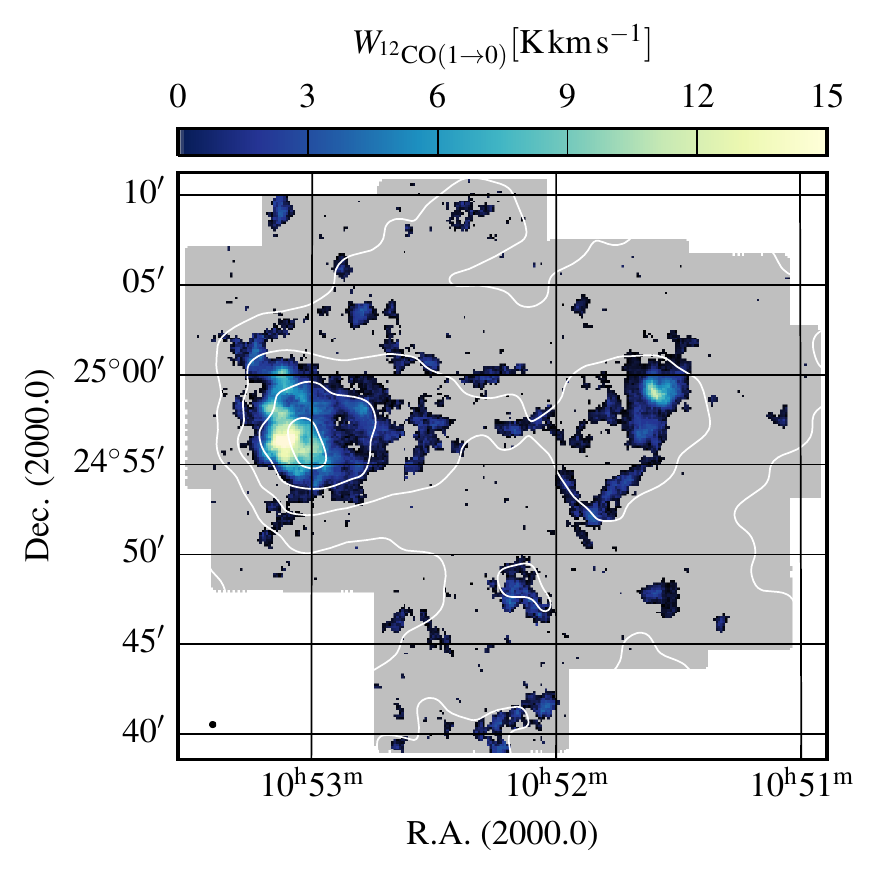}}
  \subfloat{\includegraphics[width=0.5\textwidth]{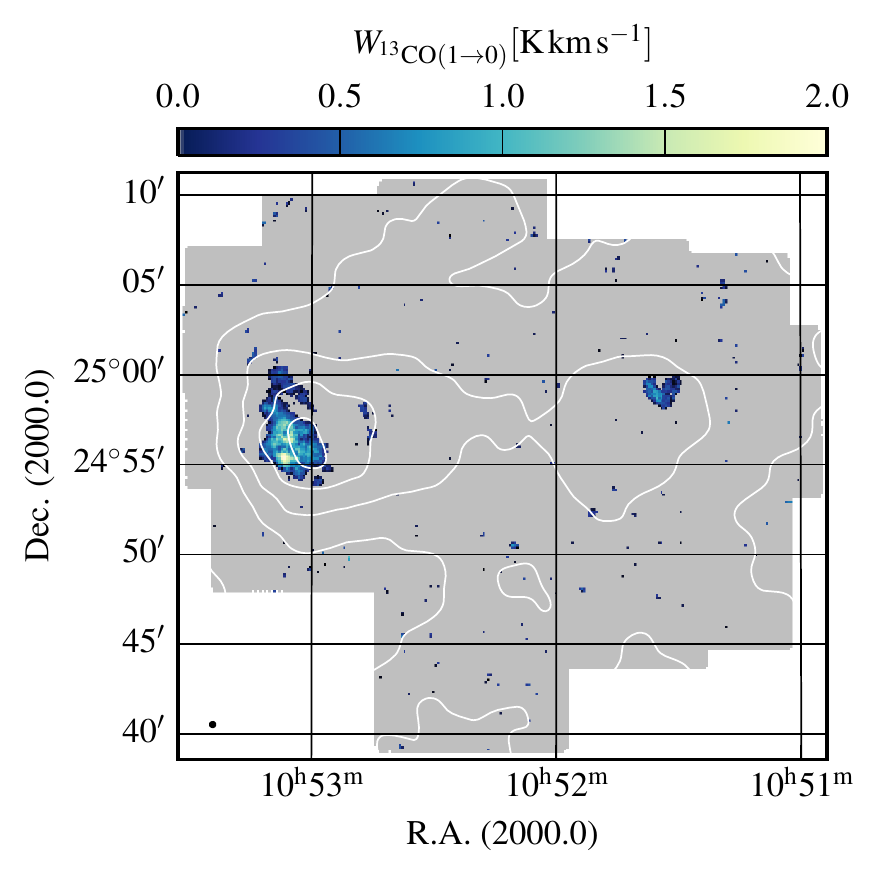}}\\
  \caption{Moment maps of the $^{12}$CO(1$\rightarrow$0) (\textit{left}) and $^{13}$CO(1$\rightarrow$0) emission (\textit{right}) of the mIVC integrated between $-45.8\,\mathrm{km}\,\mathrm{s}^{-1}\lesssim v_\mathrm{LSR} \lesssim -33.1\,\mathrm{km}\,\mathrm{s}^{-1}$. Before integration, the data cubes are masked using a spatially and spectrally smoothed version of the data cube as a mask. The contours mark the modelled dust optical depth from \textit{Planck} from $1\times10^{-6}$ to $9\times10^{-6}$ in steps of $2\times10^{-6}$. The beam-sizes are plotted at the bottom left corner.
  }
  \label{fig:mivc-co}
\end{figure*}

As expected, the mIVC is detected in $^{12}$CO(1$\rightarrow$0) and $^{13}$CO(1$\rightarrow$0). The observed brightness temperatures range up to $\sim$${7.0\,\mathrm{K}}$ for $^{12}$CO(1$\rightarrow$0) and $\sim$${1.4\,\mathrm{K}}$ for $^{13}$CO(1$\rightarrow$0) with the FTS spectrometer. Moment maps of $^{12}$CO(1$\rightarrow$0) and $^{13}$CO(1$\rightarrow$0) integrated between $-45.8\,\mathrm{km}\,\mathrm{s}^{-1}\lesssim v_\mathrm{LSR} \lesssim -33.1\,\mathrm{km}\,\mathrm{s}^{-1}$ are shown in Fig.~\ref{fig:mivc-co} superimposed by contours of dust optical depth from \textit{Planck}. {We note} that the extent of the CO maps is smaller than the previously shown maps in Fig.~\ref{fig:mivc-maps}. 

In order to emphasise the low-level CO emission, that is spread over the mIVC, the data cubes are masked using a spatially and spectrally smoothed version of the data cube as a mask. The smoothed data is clipped at $5\sigma$ of its pixel-based noise map. {Within this mask the unsmoothed CO emission is integrated.}

The CO distribution exhibits three different components (Fig.~\ref{fig:mivc-co}): Most of the $^{12}$CO and $^{13}$CO is found at the eastern edge {positionally coincident with the FIR maximum.} A second lower $^{12}$CO peak is detected near the second FIR peak to the west. In addition, multiple small CO clumps are spread across the entire mIVC, however, only detectable in $^{12}$CO(1$\rightarrow$0). Some clumps are observed close to the field edges. Hence, the $^{12}$CO is expected to extend even beyond the {mapped areas}. In $^{13}$CO only the brightest $^{12}$CO features are detected.

\begin{figure}[!t]
  \centering
  \includegraphics[width=0.5\textwidth]{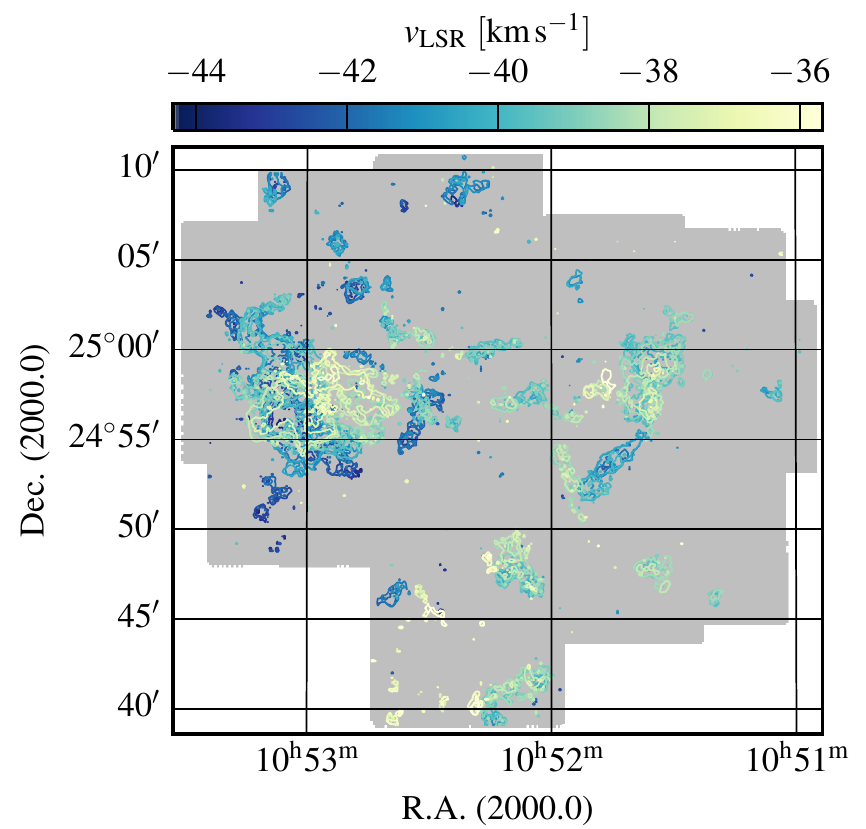}
  \caption{Renzogram of $^{12}$CO(1$\rightarrow$0) emission of the mIVC. For each spectral channel a contour at an intensity value of $0.5\,$K is plotted. The colours encode the radial velocity of the channel.
  }
  \label{fig:mivc-renzo}
\end{figure}

The velocity structure of the $^{12}$CO(1$\rightarrow$0) emission is shown as a renzogram \citep[e.g.][]{Schiminovich1997} in Fig.~\ref{fig:mivc-renzo}. In a renzogram for each channel of the masked data cube a contour at a fixed intensity is plotted. Here, we choose $0.5\,\mathrm{K}$ emphasising low-level emission. The colour of each contour corresponds to the radial velocity of the channel. Hence, a renzogram shows the structure of the {brightness temperature distribution} as the shape of the contour and the velocity by the colour {of the contour}.

The renzogram reveals the complicated velocity structure of the mIVC. There is no coherent gradient across the mIVC. In some $^{12}$CO clumps the emission is shifted significantly even between neighbouring spectral channels. 

The bright CO peaks are spread over a large range of velocities. At the eastern edge there is a CO structure between $-42\,\mathrm{km}\,\mathrm{s}^{-1} \lesssim v_\mathrm{LSR} \lesssim -40\,\mathrm{km}\,\mathrm{s}^{-1}$ that is elongated in north-south direction along the edge of the cloud towards a region of low $N_\mathrm{HI}$ \citep{Roehser2014}. A second structure at lower absolute velocities is located on top that extends further to the west. These two components in the $^{12}$CO emission have been {reported} already by \citet[][their Fig.~1b]{Desert1990a}.

There is a bimodal velocity structure within the mIVC (Fig.~\ref{fig:average-spectra}, right). By summing up all the $^{12}$CO(1$\rightarrow$0) emission in each channel separately, the distribution of the emission shows two peaks, the {brighter} one at $v_\mathrm{LSR}\simeq-38.8\,\mathrm{km}\,\mathrm{s}^{-1}$ and the second {fainter} peak at $v_\mathrm{LSR}\simeq-41.5\,\mathrm{km}\,\mathrm{s}^{-1}$. 
Many of the small molecular condensations show an elongation from bottom-left to top-right (Fig.~\ref{fig:mivc-renzo}).

\subsubsection{Deriving H$_2$ column densities}
\label{sec:mivc-h2}

\begin{figure}[!t]
  \centering
  \includegraphics[width=0.5\textwidth]{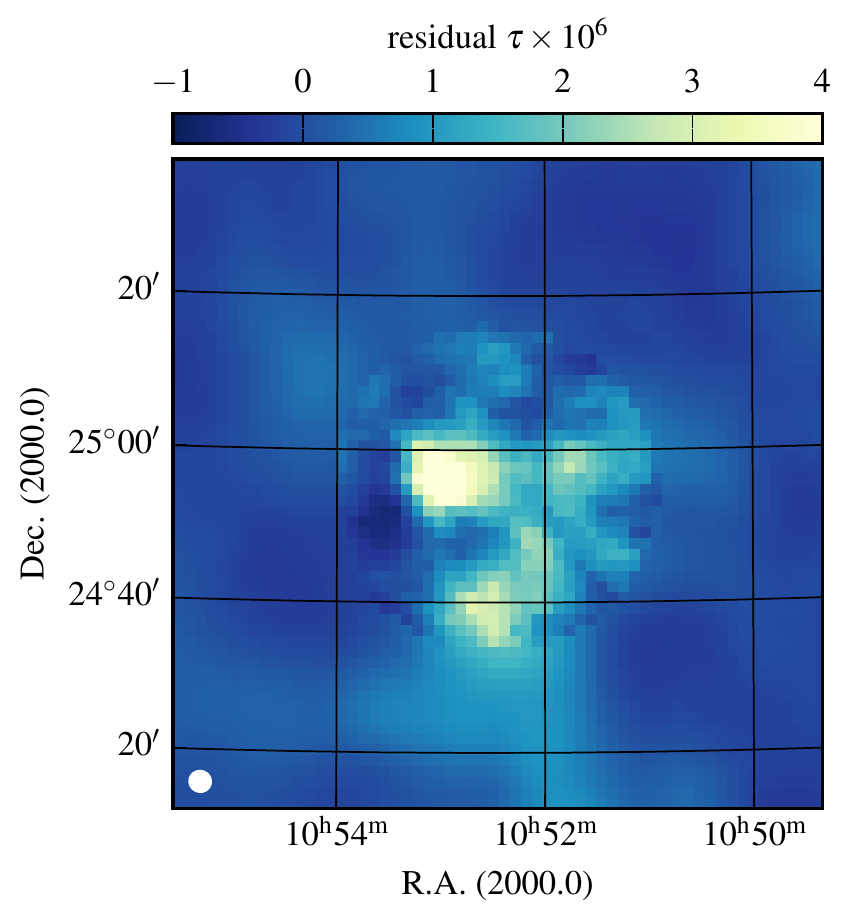}
  \caption{{Residual dust optical depth $\tau$ of the mIVC after subtracting the best model with the linear parameters given in Table \ref{tab:hi-ir-coeff}.}
  }
  \label{fig:mivc-res-tau}
\end{figure}

The FIR dust continuum emission traces the total hydrogen column density. {By fitting the linear part of the HI-$\tau$ correlation we remove the dust emission that is associated with the atomic medium. The map of the residual $\tau$ {(Fig.~\ref{fig:mivc-res-tau})}
resembles the underlying FIR emission with its double-peak structure (compare with Fig.~\ref{fig:mivc-maps}, bottom-right). By applying Eq.~\eqref{eq:nh2-ivc} and the fitted linear parameters (Table \ref{tab:hi-ir-coeff}) the distribution of H$_2$ column densities across the mIVC is calculated.}
The resulting $N_{\mathrm{H}_2}^\mathrm{IVC}$ distribution follows the FIR emission with a peak value of $N_{\mathrm{H}_2}^\mathrm{IVC}\simeq{2.7}\times10^{20}\,\mathrm{cm}^{-2}$. 

As a measure for the uncertainty of the derived $N_{\mathrm{H}_2}^\mathrm{IVC}$ map we use the scatter of the residual FIR emission, {which is} the difference between the observed and modelled $\tau$ {as defined in Eq.~\eqref{eq:h-ir-corr2}}. This scatter reflects the intrinsic uncertainties of the HI-$\tau$ correlation, which is $\Delta N_{\mathrm{H}_2}^\mathrm{IVC}\simeq0.1\times10^{20}\,\mathrm{cm}^{-2}$.

\subsubsection{The $X_\mathrm{CO}$-factor}
\label{sec:xco-factor}

{We compare the measured $^{12}$CO(1$\rightarrow$0) emission to the H$_2$ derived from \textit{Planck}} in order to infer the $X_\mathrm{CO}$ conversion factor {$X_{\mathrm{CO}} = {N_{\mathrm{H}_2}^\mathrm{IVC}}/{W_{^{12}\mathrm{CO}(1\rightarrow0)}}$ \citep[e.g.][]{Bolatto2013}} and its variations across the cloud. For the comparison, the integrated $^{12}$CO(1$\rightarrow$0) emission $W_{^{12}\mathrm{CO}(1\rightarrow0)}$ is smoothed to the spatial resolution of {\textit{Planck}}.

\begin{figure}[!t]
  \centering
  \includegraphics[width=0.5\textwidth]{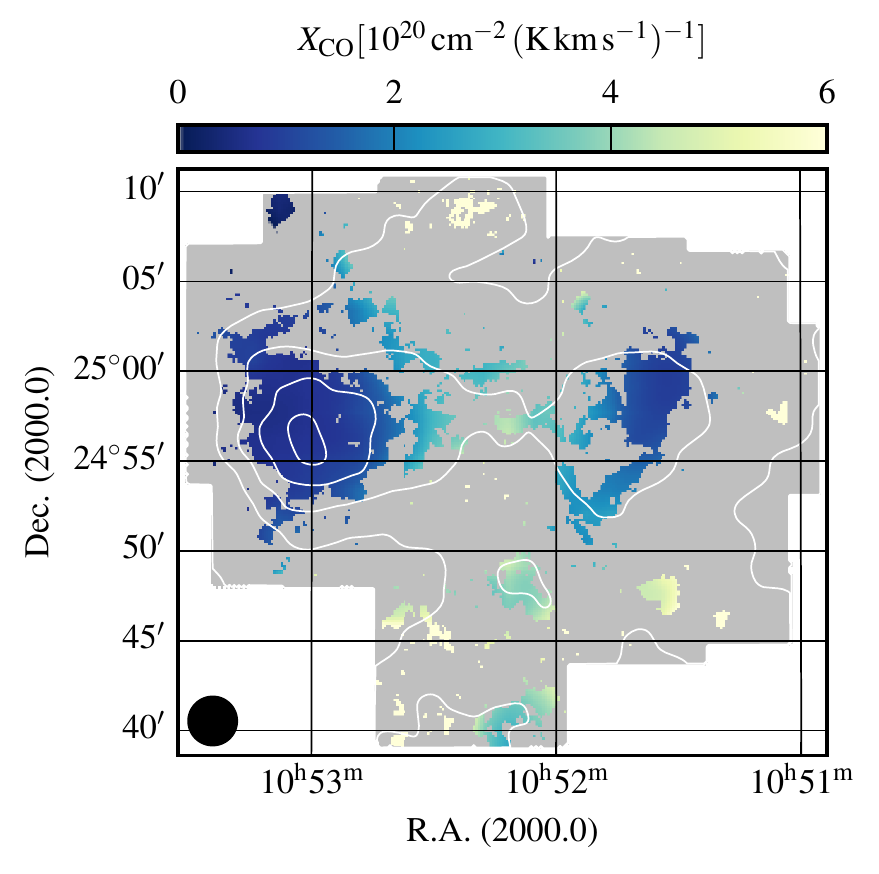}
  \caption{$X_{\mathrm{CO}}$ map of the mIVC {at the locations of detected $^{12}$CO(1$\rightarrow$0) emission}. The contours mark the modelled dust optical depth from \textit{Planck} from $1\times10^{-6}$ to $9\times10^{-6}$ in steps of $2\times10^{-6}$.
  The values range between ${0.5}\times10^{20}\,\mathrm{cm}^{-2}\,(\mathrm{K}\,\mathrm{km}\,\mathrm{s}^{-1})^{-1}\lesssim X_\mathrm{CO}\lesssim{11}\times10^{20}\,\mathrm{cm}^{-2}\,(\mathrm{K}\,\mathrm{km}\,\mathrm{s}^{-1})^{-1}$. {We note} that the $X_\mathrm{CO}$ map is derived at the angular resolution of the \textit{Planck} data, which is shown at the bottom-left.}
  \label{fig:mivc-xco}
\end{figure}

Figure \ref{fig:mivc-xco} shows the resulting $X_\mathrm{CO}$ map of the mIVC superimposed with the distribution of the dust optical depth as contours.
The values range between ${0.5}\times10^{20}\,\mathrm{cm}^{-2}\,(\mathrm{K}\,\mathrm{km}\,\mathrm{s}^{-1})^{-1}\lesssim X_\mathrm{CO}\lesssim{11}\times10^{20}\,\mathrm{cm}^{-2}\,(\mathrm{K}\,\mathrm{km}\,\mathrm{s}^{-1})^{-1}$ spreading significantly around the canonical $2\times10^{20}\,\mathrm{cm}^{-2}\,(\mathrm{K}\,\mathrm{km}\,\mathrm{s}^{-1})^{-1}$ \citep{Bolatto2013}. 

We derive a mean $X_\mathrm{CO}$ value by the comparison of the total $N_{\mathrm{H}_2}^\mathrm{IVC}$ column density and the total $^{12}$CO(1$\rightarrow$0) flux across the mIVC. All negative values in the $N_{\mathrm{H}_2}^\mathrm{IVC}$ map are blanked. This approach yields $\bar{X}_\mathrm{CO}\simeq{1.8}\times10^{20}\,\mathrm{cm}^{-2}\,(\mathrm{K}\,\mathrm{km}\,\mathrm{s}^{-1})^{-1}$ across the entire cloud. The inferred mean $\bar{X}_\mathrm{CO}$ value is consistent with the canonical $\sim$$2\times10^{20}\,\mathrm{cm}^{-2}\,(\mathrm{K}\,\mathrm{km}\,\mathrm{s}^{-1})^{-1}$ for the Milky Way \citep{Bolatto2013}.

\subsubsection{Excitation conditions of CO}
\label{sec:excitation-co}

We analyse the excitation conditions of CO by using the standard formulas as given by \citet{Pineda2008} that assume local thermal equilibrium (LTE). This allows us to estimate the excitation temperature $T_\mathrm{ex}$, the optical depth $\tau_{^{13}\mathrm{CO}}$, and the column density $N_{^{13}\mathrm{CO}}$. $^{12}$CO gets optically thick easily which does not allow a meaningful calculation of the same quantities for $^{12}$CO.

\begin{figure*}[!t]
  \centering
  \subfloat{\includegraphics[width=0.45\textwidth]{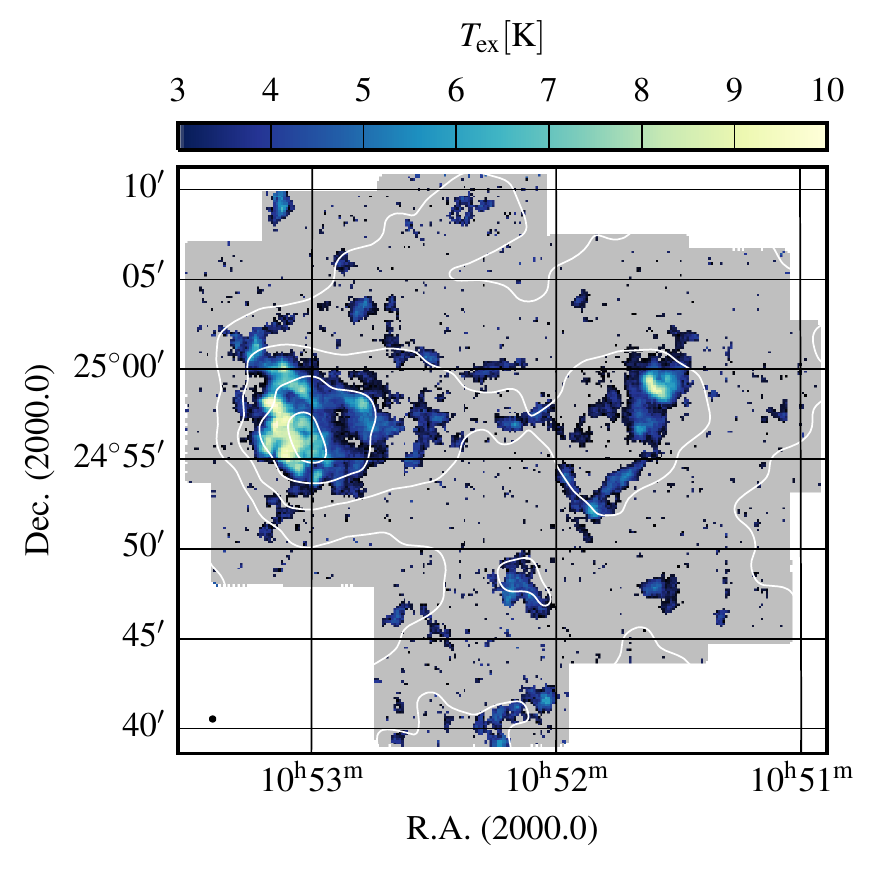}}
  \hspace{0.5cm}
  \subfloat{\includegraphics[width=0.45\textwidth]{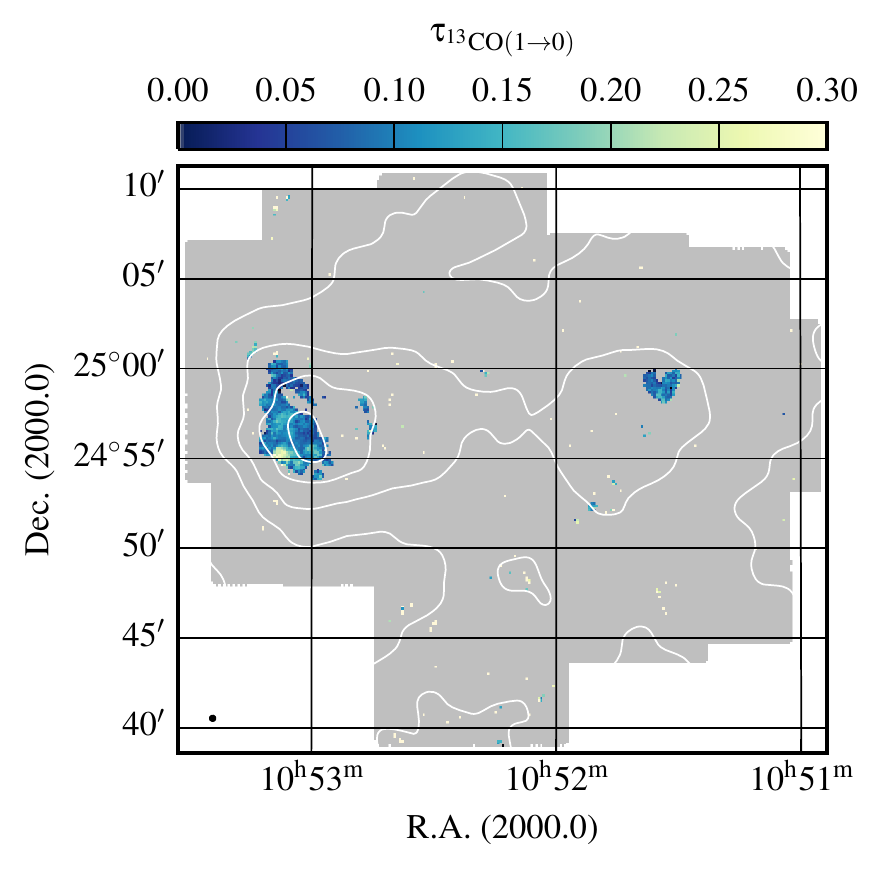}}\\
  \subfloat{\includegraphics[width=0.45\textwidth]{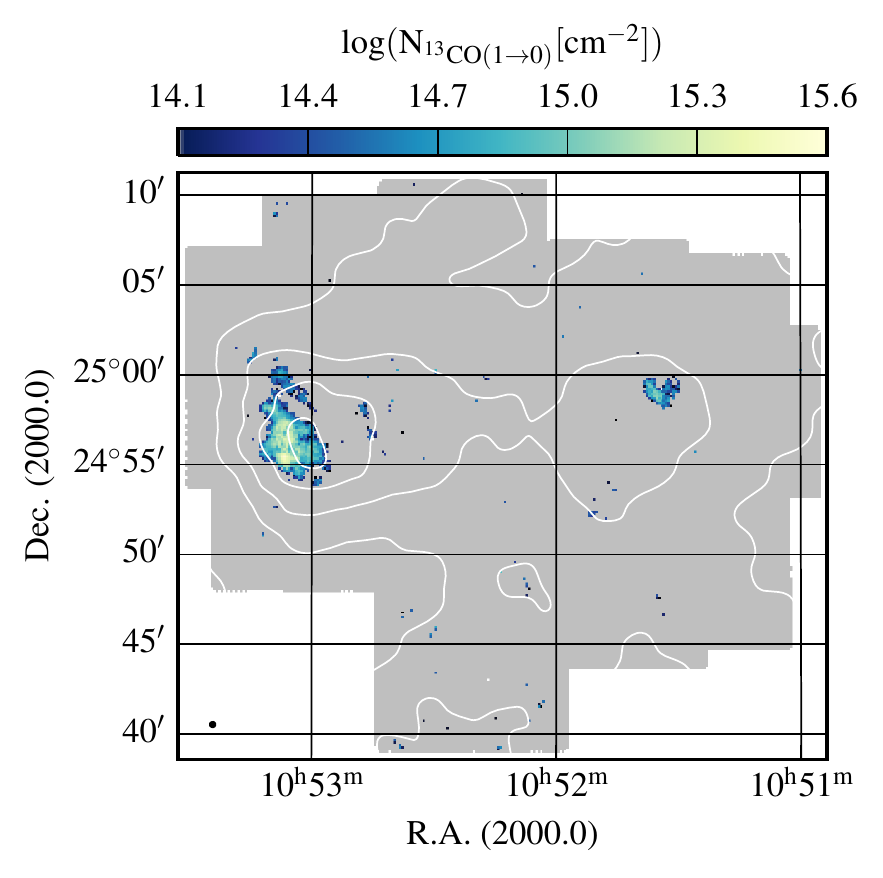}}
  \hspace{0.5cm}
  \subfloat{\includegraphics[width=0.45\textwidth]{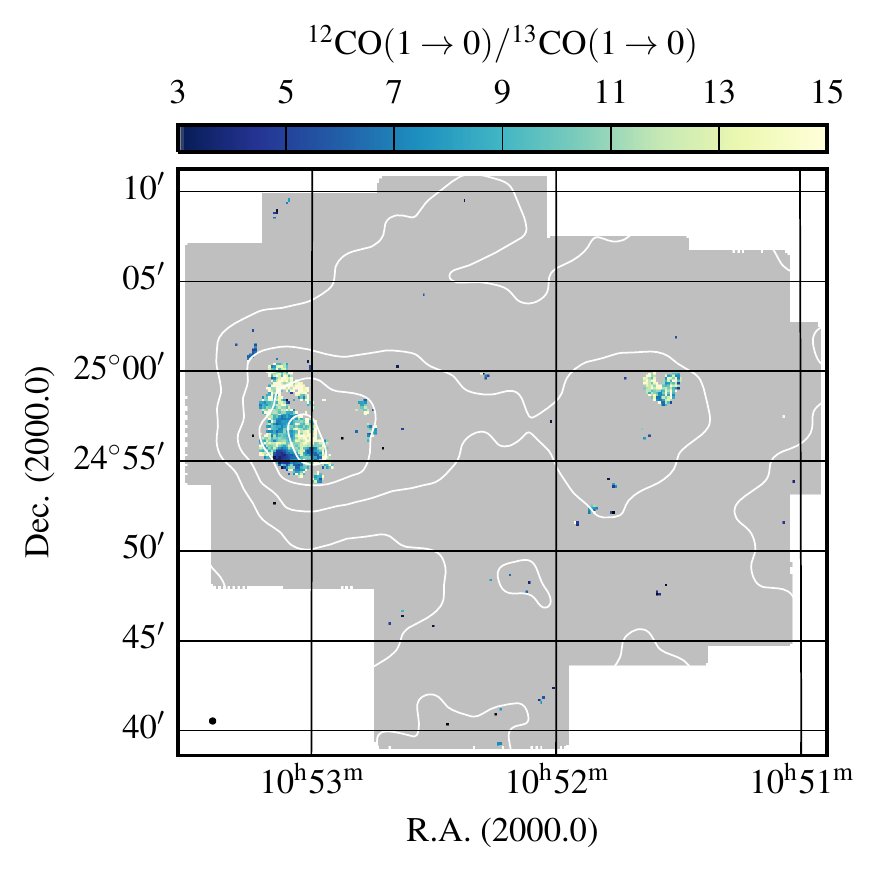}}\\
  \caption{The plots show the excitation temperature $T_\mathrm{ex}$ (\textit{top-left}), the optical depth $\tau_{^{13}\mathrm{CO}}$ (\textit{top-right}), the column density $N_{^{13}\mathrm{CO}}$ (\textit{bottom-left}), and the peak intensity ratios $^{12}$CO/$^{13}$CO (\textit{bottom-right}) of the mIVC. The contours mark the dust optical depth from \textit{Planck} from $1\times10^{-6}$ to $9\times10^{-6}$ in steps of $2\times10^{-6}$. At the bottom left the beam size of the IRAM $^{13}$CO(1$\rightarrow$0) data is plotted.
  }
  \label{fig:mivc-excitation}
\end{figure*}

The maps of $T_\mathrm{ex}$, $\tau_{^{13}\mathrm{CO}}$, $N_{^{13}\mathrm{CO}}$, and $^{12}$CO/$^{13}$CO are shown in Fig.~\ref{fig:mivc-excitation}. The excitation temperatures are $3.5\,\mathrm{K}\lesssim T_\mathrm{ex} \lesssim 10.5\,\mathrm{K}$, significantly lower than the dust temperature with $T_\mathrm{d}\gtrsim18.8\,\mathrm{K}$ \citep{Planckcollaboration2014XI}. The derived optical depth are $\tau_{^{13}\mathrm{CO}}\lesssim0.3$ and the $^{13}$CO column densities range up to $N_{^{13}\mathrm{CO}} \lesssim3.3\times10^{15}\,\mathrm{cm}^{-2}$. The ratios $^{12}$CO/$^{13}$CO are $4-20$. Generally, these values are close to those of giant molecular clouds in the Milky Way \citep[e.g.][]{Pineda2008,Pineda2010}.

One particular feature at the eastern edge of the mIVC is noticeable bright in $^{13}$CO as indicated by the largest optical depth and lowest $^{12}$CO/$^{13}$CO intensity ratios. Most likely this may correspond to the largest CO column densities, which are located at the very edge of the CO distribution, suggesting a very steep column density gradient.

\subsubsection{Radiative transfer calculations}
\label{sec:radiative-transfer}

A detailed modelling of the radiative transfer within the mIVC is beyond the scope of this paper. In order to obtain an estimate on the kinetic temperature and density within the mIVC, we use the publicly available statistical equilibrium radiative transfer code RADEX \citep{vanderTak2007} to approximate the observed and inferred parameters towards the brightest $^{13}$CO(1$\rightarrow$0) emission. We use a grid of different input values for the kinetic temperature, the H$_2$ density, and the line width. The parameters that match the observed and inferred quantities best are $T_\mathrm{kin}\simeq52\,\mathrm{K}$ and $n_{\mathrm{H}_2}\simeq440\,\mathrm{cm}^{-3}$. The density is below that of the critical density of the line transition of $1.9\times10^3\,\mathrm{cm}^{-3}$ for $^{13}$CO(1$\rightarrow$0) \citep[e.g.][]{Hernandez2011}. Similarly, the inferred kinetic temperature is much higher then the derived excitation temperatures, indicating sub-thermal excitation conditions within the mIVC.

\subsubsection{Comparing HI and CO}
\label{sec:mivc-hi-co}

In the following we compare the distribution of WSRT HI and IRAM $^{12}$CO(1$\rightarrow$0) emission in the mIVC in more detail. The WSRT data is smoothed to a circular beam of $49.1\arcsec$, {to which the IRAM $^{12}$CO(1$\rightarrow$0) data is smoothed as well}. In addition, the $^{12}$CO data is re-gridded and smoothed in velocity to the spectral resolution of the WSRT observations. 

{In Fig.~\ref{fig:average-spectra} (right) we already showed the average HI and $^{12}$CO(1$\rightarrow$0) spectra for the mIVC revealing evidence for systemic velocity shifts between the individual EBHIS, WSRT, and IRAM data sets.}

In Fig.~\ref{fig:mivc-spatial-cuts} {a spatial slice} through the integrated HI and $^{12}$CO maps is shown {for the mIVC} towards lower right ascension (from left to right) {for $24^\circ 50' \leq \mathrm{Dec.}\leq 25^\circ 5'$ (compare with Fig.~\ref{fig:mivc-co}, left). The median profile is calculated for the masked $^{12}$CO data.} The fluxes are normalised to the maximum value of the slice. The integrated HI is given by the black dashed line, the integrated $^{12}$CO(1$\rightarrow$0) by the red solid line.
Figure \ref{fig:mivc-spatial-cuts} illustrates several things about the distribution of HI and CO:
\begin{itemize}
 \item The CO is embedded within the HI.
 \item {The brightest $^{12}$CO peak has} a significant spatial offset to the WSRT HI maximum at the eastern edge. The CO is located towards the rim of the mIVC.
 \item {While at the eastern edge the relative amounts of HI and CO are similar, towards the centre of the mIVC the relative fraction of CO decreases.}
 \item {The HI emission at the eastern edge is angularly less extended than at the cloud centre.}
 \item Along the spatial slice {many of the individual HI peaks can be associated with a corresponding CO peak. However, there is often a displacement between the HI and CO.}
\end{itemize}

Under equilibrium conditions it is thought that the spatial location of HI and CO peaks should coincide. The atomic hydrogen turns molecular first, followed by the transition from ionised carbon to CO at larger densities \citep[e.g.][]{Snow2006}. Thus, there is a positionally tighter connection between H$_2$ and CO rather than between HI and CO.

\subsection{The spatial structure of the atomic and molecular IVC}
\label{sec:comparison}

{We measure for the mIVC $\sim$82\% of the total WSRT HI flux within the aIVC mosaic. The corresponding EBHIS HI flux of the mIVC is $\sim$38\% of that of the aIVC. Hence, for the mIVC more HI flux is recovered by the interferometer relative to the total amount of single-dish HI gas. This suggests that on angular scales below the angular resolution of EBHIS the mIVC contains more power than the aIVC.}

A quantitative measure for the structure of a cloud is the azimuthally averaged power spectral density (PSD). The PSD is calculated as the square of the modulus of the 2D Fourier transform of the integrated maps. By averaging the PSD {at constant wavenumber $k$}, a radial profile is obtained. {Before the Fourier transform is computed, the images are apodised at the edges of the interferometric pointings by a Gaussian filter. The apodisation reduces effects from the Fourier transform due to artificial edges in the images \citep[e.g.][]{MivilleDeschenes2002b}.}

In Fig.~\ref{fig:psd-ivc1-ivc2} we show the resulting PSDs {separately for the EBHIS, WSRT HI data, and FIR \textit{Planck} data} of the aIVC and the mIVC. {Each data set is analysed only within the region covered by the WSRT primary beam (compare with Figs.~\ref{fig:aivc-maps} and \ref{fig:mivc-maps}). The two WSRT pointings for the aIVC are considered separately as aIVC1 and aIVC2 in order to achieve the same circular shape and size of the different interferometric observations. The different colours correspond either to EBHIS, WSRT, or Planck, and mIVC, aIVC1, or aIVC2. The angular resolutions of the different observations are used (Table \ref{tab:data}) because any smoothing may shift power between scales. The angular resolution limits are marked by the vertical black lines. These are labelled at the top with the corresponding angular scale of the (major axes) of the beams.

Because of the elliptical interferometric beam, the 2D power spectra are deconvolved in the Fourier domain by the division of the Fourier transform of the respective beam. After this deconvolution the 1D PSDs are calculated. In Fig.~\ref{fig:psd-ivc1-ivc2} these PSDs are plotted as the thick lines. The thin lines show the PSDs without beam deconvolution. The beam-deconvolved EBHIS and \textit{Planck} data is plotted within the full data range while the original un-deconvolved data is shown up to spatial frequencies of $1/10.8\arcmin$ and $1/5.27\arcmin$, respectively, for visualisation purposes.

The EBHIS data is normalised to the total power within the respective pointing. Similarly, the individual WSRT pointings are normalised to the corresponding single dish flux. The \textit{Planck} PSDs are shifted to the bottom in order to avoid confusion with the other profiles. Hence, the \textit{Planck} profiles are arbitrarily scaled.

In order to estimate at what angular scales the PSDs of the HI column density maps are dominated by the noise, we compare PSDs of individual channels without beam deconvolution in channels of emission and noise. We find that at angular scales larger than the beam major axis of the WSRT data the PSDs are dominated by the HI emission and not by the noise.}

\begin{figure}[!t]
  \centering
  \includegraphics[width=0.45\textwidth]{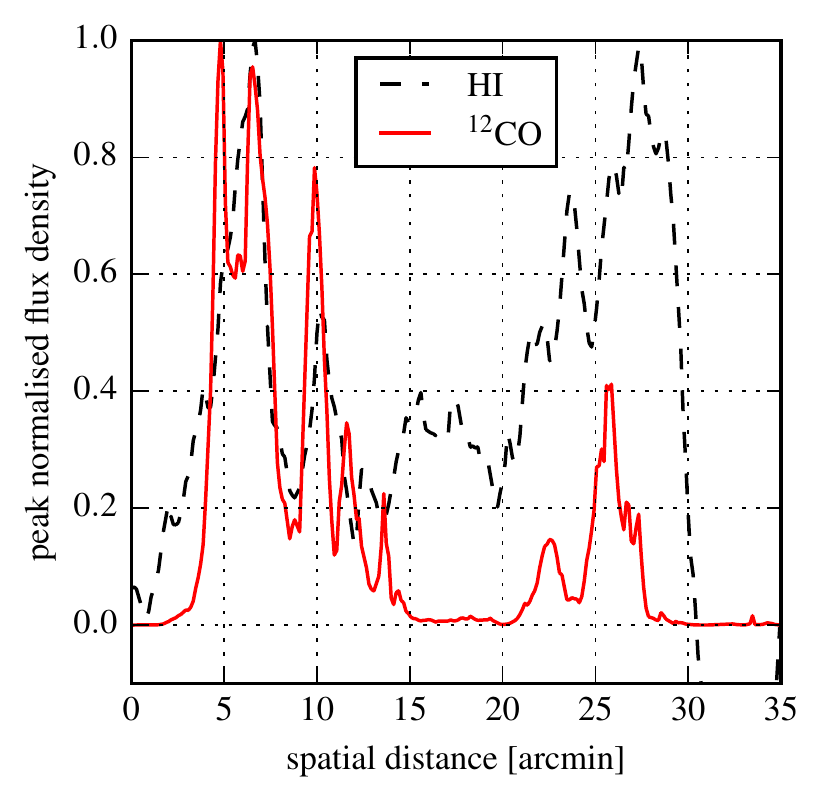}
  \caption{{Median} spatial slice across the integrated HI (black-dashed) and $^{12}$CO(1$\rightarrow$0) (red-solid) map. The slice is calculated towards lower Right Ascension {for $24^\circ 50' \leq \mathrm{Dec.}\leq 25^\circ 5'$.} The integrated flux is normalised to the maximum along the slice.
  }
  \label{fig:mivc-spatial-cuts}
\end{figure}

\begin{figure}[!t]
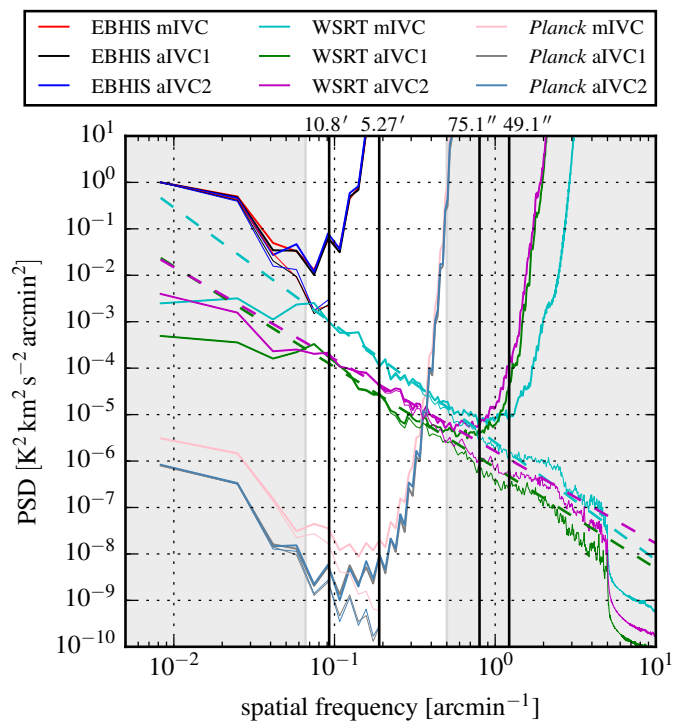

  \centering
  \includegraphics[width=0.5\textwidth]{{{spectral_power_1_2a_2b_unsmoothed_single_apodise_beam_normalised_Planck}}}
  \caption{Azimuthally averaged power spectral densities (PSDs) of the aIVC and mIVC. {The thick lines show the power spectra with beam deconvolution, the thin lines without. Within the white area power laws (the dashed lines) are fitted to the deconvolved WSRT data.}
  }
  \label{fig:psd-ivc1-ivc2}
\end{figure}

{The PSDs calculated for the EBHIS data are very similar for all pointings. The normalised total fluxes of the WSRT observations are slightly different for the individual pointings. The noise of the mIVC observation is higher than for the aIVC. However, apparently the WSRT detects more power on most of the spatial scales for the mIVC than for the two aIVC pointings. This excess of power appears to be related to an offset in amplitude of the mIVC profile as is seen from the un-deconvolved data. {This different amplitude is likely related to the higher HI column densities of the mIVC \citep{Gautier1992,MivilleDeschenes2007}. For both clouds the amplitude is expected to scale proportional to the column density as $N_\mathrm{HI}^{2.0\pm0.1}$ \citep{MivilleDeschenes2007}. Thus, a factor of up to {approximately six} in amplitude is expected, which is close to the difference in the PSDs.}

The dust optical depths from \textit{Planck} provide a measure of the total amount of gas and dust within the clouds. Consequently, the PSDs for the \textit{Planck} data show that the mIVC contains more power than the aIVC on all scales.

{The coloured dashed lines in Fig.~\ref{fig:psd-ivc1-ivc2}} represent fits of power-laws to the {beam-deconvolved} WSRT PSD profiles. {These power laws are fitted within the white region. The lower bound corresponds to the largest angular scale to which the WSRT is sensitive, corresponding to the shortest baseline of 48\,m, equivalent to $\sim$$15\arcmin$. The upper bound is set below the increase of power due to the beam-deconvolution. Within this range a single power law is a good approximation for all three PSDs. The fitted slopes are $-2.52\pm0.09$ for the mIVC, $-2.17\pm0.11$ and $-1.98\pm0.07$ for the aIVC1 and aIVC2 pointings.} {These slopes, especially those for the aIVC, are somewhat shallower than what is typically found in other studies} \citep[e.g.][]{MivilleDeschenes2007,Burkhart2013b,Roy2015}. {The slopes of the aIVC may be affected slightly by the noise in the high-resolution HI data, although the power-law is fitted in a range of angular scales that have a high signal-to-noise.}

\section{Discussion}
\label{sec:discussion}

\subsection{The molecular IVC}
\label{sec:disc-mivc}

\subsubsection{The $X_\mathrm{CO}$ factor}
\label{sec:disc-mivc-xco}

The quantitative comparison of H$_2$ column densities and $^{12}$CO(1$\rightarrow$0) emission results into conversion factors ${0.5}\times10^{20}\,\mathrm{cm}^{-2}\,(\mathrm{K}\,\mathrm{km}\,\mathrm{s}^{-1})^{-1}\lesssim X_\mathrm{CO}\lesssim{11}\times10^{20}\,\mathrm{cm}^{-2}\,(\mathrm{K}\,\mathrm{km}\,\mathrm{s}^{-1})^{-1}$ (Section  \ref{sec:xco-factor}). {The mean conversion factor $\bar{X}_\mathrm{CO}\simeq{1.8}\times10^{20}\,\mathrm{cm}^{-2}\,(\mathrm{K}\,\mathrm{km}\,\mathrm{s}^{-1})^{-1}$ across the entire mIVC is consistent with the canonical value \citep{Bolatto2013}.} The lowest conversion factors are derived at the locations of the CO peaks. Thus, CO-darkish H$_2$ gas \citep{Grenier2005,Wolfire2010,Planckcollaboration2011XIX} is found mostly in regions of low CO abundances where $X_\mathrm{CO}$ is strongly enhanced.

{Several effects contribute to apparent changes in the $X_\mathrm{CO}$ factors that cannot be disentangled here. Firstly, there are real variations of the conversion factor due to different relative amounts of H$_2$ and CO across the cloud. Secondly, there is CO-dark H$_2$ gas mostly towards lower column densities \citep{Wolfire2010}. Thirdly, changes in the dust properties may change the HI-FIR correlation and the inferred amount of H$_2$. {Such changes are not only inferred at high column densities \citep[e.g.][]{Ormel2011}. Recent studies \citep{Ysard2015,Fanciullo2015} show that even within the diffuse ISM, at HI column densities that we probe here, the dust emission properties vary. However, these variations cannot account for FIR excess in total because they are too small.}

Studies of high-latitude IVMCs have in common that the authors attribute bright CO emission and large molecular abundances in general to dynamical phenomena in the Galactic halo \citep[e.g.][]{Herbstmeier1993,Herbstmeier1994,Moritz1998,Weiss1999,Lenz2015}. However, according to \citet{Herbstmeier1994} the excitation conditions of CO, {for example} line ratios, are similar to other molecular clouds. They propose that either the CO abundances are unusual high or that CO is more efficiently excited.

\subsubsection{Excitation conditions of CO}
\label{sec:disc-excitation}

For the mIVC we obtain similar values for the excitation temperature $T_\mathrm{ex}$, optical depth $\tau_{^{13}\mathrm{CO}(1\rightarrow0)}$, and $^{13}$CO column density $N_{^{13}\mathrm{CO}}$ as {for example} \citet{Pineda2008,Pineda2010}. For Perseus \citet{Pineda2008} find that about 60\% of the $^{12}$CO(1$\rightarrow$0) emission is sub-thermally excited corresponding to volume densities below $1\times10^3\,\mathrm{cm}^{-3}$ \citep[e.g.][]{Snow2006}. Sub-thermal excitation may be also very important for the mIVC since the RADEX grid calculations reproduce the observed peak $^{13}$CO(1$\rightarrow$0) emission best for $T_\mathrm{kin}\simeq45\,\mathrm{K}$ and $n_{\mathrm{H}_2}\simeq440\,\mathrm{cm}^{-3}$ (Section \ref{sec:radiative-transfer}) which is significantly below the critical density of $^{12}$CO(1$\rightarrow$0) or $^{13}$CO(1$\rightarrow$0). 

According to \citet{Liszt2010}, the specific brightness $W_\mathrm{CO}/N_\mathrm{CO}$ is larger in warm and sub-thermally excited gas. Such environments correspond to kinetic temperatures that are much higher than the CO(1$\rightarrow$0) excitation temperature, in agreement with {our findings}. Generally, CO chemistry is more sensitive to the environmental conditions than H$_2$ is \citep[e.g.][]{Liszt2012}. For some of their lines-of-sight \citet{Liszt2012} describe strongly over-pressured molecular clumps which are likely transient.

\subsubsection{Evidence for non-equilibrium conditions}
\label{sec:disc-mivc-non-equi}

There are observational indications that the formation of H$_2$ and CO in the mIVC does not occur in formation-dissociation equilibrium:
\begin{itemize}
 \item There are spatial {displacements} between all the different data sets (Figs.~\ref{fig:mivc-maps} and \ref{fig:mivc-spatial-cuts}). Also the {velocity} distribution of the HI and CO emission is different (Fig.~\ref{fig:average-spectra}), {suggesting varying H$_2$ formation efficiencies for different spectral components}.
 \item The brightest CO emission is found at the edge of the integrated CO emission map, located eastwards of the nearby HI maximum close to the cloud's rim (Fig.~\ref{fig:mivc-spatial-cuts}). The $^{12}$CO(1$\rightarrow$0)/$^{13}$CO(1$\rightarrow$0) ratios are lowest at one particular spot at the eastern side (Fig.~\ref{fig:mivc-excitation}) probably linked to the largest molecular abundances and column density contrasts. {This is reminiscent of the Draco molecular cloud \citep[e.g.][]{MivilleDeschenes2016}.}
 \item In general, the spectral and spatial properties of the mIVC are complicated. We observe several velocity components in both HI and CO and a rich clumpy structure (Figs.~\ref{fig:mivc-maps}, \ref{fig:mivc-renzo}). In $^{12}$CO(1$\rightarrow$0), there is a bimodal velocity distribution across the cloud but no coherent velocity gradient.
 \item The observed radial velocity of the mIVC is $\sim$$-40\,$km\,s$^{-1}$ moving through a thin halo medium. This situation in itself may be unstable and subject to instabilities {\citep[e.g.][]{MivilleDeschenes2016}.}
\end{itemize}

\subsubsection{Is the mIVC able to form stars?}
\label{sec:star-formation}

All known star-forming high-latitude clouds are not classified as IVCs \citep{McGehee2008}. Thus, it would be surprising to find evidences of star formation in the mIVC. Within the innermost part of the mIVC, that we measured with IRAM, the total combined single-dish HI and H$_2$ mass is about $M_\mathrm{H}\simeq{42}\,M_\odot$. This mass is low compared to other star-forming high-latitude clouds {like MBM\,12 \citep{Pound1990,Luhman2001} or MBM\,20 \citep{Liljestrom1991,Hearty2000}.} However, the mIVC is connected to more extended HI structures that contain significantly more mass \citep{Roehser2014}.

Using the virial parameter \citep{Bertoldi1992}
\begin{equation}
 \alpha_\mathrm{vir}=\frac{M_\mathrm{vir}}{M} = \frac{5\sigma^2R}{GM} \simeq 1.2 \left( \frac{\sigma_\mathrm{v}}{\mathrm{km}\,\mathrm{s}^{-1}} \right)^2 \left( \frac{R}{\mathrm{pc}} \right) \left( \frac{M}{10^3\,\mathrm{M}_\odot} \right)^{-1},
\end{equation}
we estimate the importance of self-gravity for the individual clumps. For clumps with $\alpha_\mathrm{vir}>>1$ gravity is unimportant, while for $\alpha_\mathrm{vir}\simeq1$ the gravitational energy is comparable to the kinetic energy. We calculate the size of each clump by summing over the number of its individual pixels and converting it to the radius of a sphere with the same angular size. Using the estimated radius and gas mass for a typical line-width of CO of $\Delta v \simeq 1\,\mathrm{km}\,\mathrm{s}^{-1}$, the typical virial parameters are $\alpha_\mathrm{vir}\simeq{120}$ with a minimum value of approximately six. Hence, the combined atomic and molecular gas within the spatial regions of the CO clumps is unlikely to be gravitationally bound.

The result is similar when we consider the cloud globally. For the entire region covered with IRAM we get $\alpha_\mathrm{virial}\simeq6$. Thus, also globally the cloud seems to be gravitationally unbound. The mIVC appears not to form stars which is expected because the densities and masses are not sufficiently large to form gravitationally bound structures.

\subsection{The formation of molecular clouds}
\label{sec:disc-formation-molecular-clouds}

Turbulence is thought to be important for the formation of molecular clouds and for subsequent star formation \citep{MacLow2004}. Simulations of turbulent colliding flows of initially warm neutral medium (WNM) show the formation of non-linear density perturbations that lead to gaseous structures of cold neutral medium (CNM) due to dynamical and thermal instabilities \citep[e.g.][]{Audit2005,Heitsch2005,Glover2007}. The H$_2$ formation is rapid with time-scales of $\sim$$1\,$Myr or less.

The main limiting factor for H$_2$ formation is the time-dependent column density distribution which continuously re-exposes the molecular material to the radiation field \citep{Heitsch2006a}. In their Fig.~7 \citet{Glover2007} show that the chemical abundances are not in equilibrium indicating that molecular cores are likely transient features \citep{VazquezSemadeni2005}. 

{We infer comparably shallow slopes of the PSDs computed for the interferomteric HI data. {Possible explanations are} high Mach numbers, which appear to flatten the PSD profiles \citep{Burkhart2013b}, {and the thermal instability of atomic gas \citep{Field1965}}. Molecules form in small and dense environments, which suggests a flattening of the PSD for smaller spatial scales of a molecular cloud. However, a steeper PSD is inferred for the molecular cloud than for the atomic one. This may be interesting for the mechanisms that form structures in general.}

{The considered single dish and interferometric HI fluxes reveal that there is apparently more atomic gas retrieved by the interferometer for the mIVC than for the aIVC as compared to the total amount of gas. {Accordingly, the HI column densities are higher within the mIVC. These may account for the different amplitudes of the PSD profiles \citep{Gautier1992,MivilleDeschenes2007}}. Apparently, this additional atomic gas is in the form of CNM rather than WNM, since the EBHIS peak spectra directly show that the mIVC contains more cold atomic gas than the aIVC \citep[][their Fig.~8]{Roehser2014}. The retrieved HI fluxes and the PSD profiles suggest the same finding but from the spatial distribution of the HI gas. 

Similarly, these additional CNM structures appear to be connected to brighter FIR emission in the mIVC, which reflects the presence of molecular hydrogen. Hence, we may have found an observational connection between CNM and molecular gas.

Nevertheless, one may argue that there should be more pronounced differences in the atomic structure of the two IVCs. These may be revealed better in proper interferometric mosaics of the targets. {Otherwise, this implies that CNM and diffuse H$_2$ are rather similar.}}

High-latitude clouds and IVMCs may be thought of as flows of warm and cold gas through the surrounding halo medium {\citep[also][]{MivilleDeschenes2016}}. Thus, it is just a consequent step to assume that, given time, such objects develop cold small-scale structures in which molecular hydrogen can form. The bimodal velocity distribution within the mIVC can be thought of as an imprint of flows of different gas components.

The motion of halo clouds through the ambient medium creates ram pressure. In the mIVC most of the H$_2$ and CO are found at the eastern side of the cloud where a sharp column density contrast at the cloud's rim is evident. We propose that the cloud moves in this direction and the largest molecular abundances are located at the leading front of the cloud {(compare also with Fig.~\ref{fig:mivc-spatial-cuts})}. Hence, ram pressure appears to accumulate gas and facilitate the formation of small-scale structures and molecules. {We note} that the unknown tangential velocity component of the cloud is likely substantial if the IVC originates from a Galactic fountain process \citep[e.g.][]{Melioli2008}.

For typical densities of the WNM \citet{Saury2014} find similarly that turbulent motions of the neutral gas alone do not cause the transition from WNM to CNM. Instead, an increase of the WNM density is required in the first place to trigger the rapid formation of CNM structures out of the WNM by turbulence. Thus for the aIVC and mIVC, ram pressure appears to be responsible for {over-pressuring} the WNM, pushing the gas to the thermally unstable regime from which the CNM is formed.

The general conditions are identical in both the aIVC and mIVC: The observed radial velocity is the same, there is cold gas with $\mathrm{FWHM}\simeq3\,\mathrm{km}\,\mathrm{s}^{-1}$, the total HI mass is even larger for the aIVC and substructure has evolved. Thus, we confirm the conclusion of \citet{Roehser2014} that the aIVC should evolve into a similar molecular IVC as the mIVC. This transition can occur rapidly, possibly within 1\,Myr \citep{Saury2014}.

In simulations of the turbulent ISM no particular triggering mechanism is required but gradually structures emerge. This is in contrast to possible interactions between IVCs and other halo clouds as the reason for the formation of molecules \citep{Herbstmeier1993,Weiss1999,Lenz2015}.

\section{Summary}
\label{sec:summary}

We present high-resolution WSRT HI and IRAM CO observations of two high-latitude intermediate-velocity clouds (IVCs). These are studied in the context of the transition from atomic to molecular clouds at the disk-halo interface. Our analysis elaborates on \citet{Roehser2014} who compared the two IVCs by using the most recent large-scale surveys in HI and the FIR, EBHIS and \textit{Planck}.

The molecular IVC (mIVC) exhibits a pronounced structure consisting of many clumps in HI and CO. This clumpy {high-column density substructure on sub-parsec scales} provides the shielding of molecules like H$_2$ and CO. Across those parts that are surveyed with IRAM, CO emission is detected indicating that the whole cloud is condensed to allow locally the formation of CO.

{Statistically, there is only weak evidence that the small-scale structures within the atomic IVC (aIVC) are different from the mIVC. In terms of HI column density, the interferometric observations detect less HI in clumps but more in a diffuse and smooth distribution.} Consequently, no CO emission is detected near the largest HI column densities of the aIVC. {The excess of interferometric HI flux for the mIVC relative to the aIVC may be considered as evidence for the larger abundance of CNM from its spatial distribution, which is reflected by the large amount of molecular gas within the mIVC. 

The estimated slopes of the PSD profiles are shallower for the atomic cloud, which is opposite to the naive expectation that molecular clouds have more substructure. This may be connected to the formation mechanisms of molecular gas in general {or to noise in our high-resolution HI data}.}

Using the dust optical depth from \textit{Planck}, we infer the column densities of molecular hydrogen within the mIVC. The $X_\mathrm{CO}$ conversion factor varies significantly across the cloud with an average $\bar{X}_\mathrm{CO}\simeq{1.8}\times10^{20}\,\mathrm{cm}^{-2}\,(\mathrm{K}\,\mathrm{km}\,\mathrm{s}^{-1})^{-1}$. The lowest $X_\mathrm{CO}$ are found at the FIR peaks increasing outwards. Thus, most of the CO-darkish H$_2$ gas is found in regions of low CO abundances.

A thorough study of similarities and differences between all high-latitude molecular and non-molecular IVCs would shed more light on a possible triggering mechanism and the requirements for the formation of such objects. It would give important insights into the Galactic fountain cycle, the fate of in-falling material, and into the evolution of the Milky Way as a whole.  

{One may have anticipated statistically more pronounced differences between the two clouds since a molecular cloud is expected to exhibit compact small-scale structures in which molecules have formed. {The 1D power spectral densities do not reveal significant differences between both clouds. This may be interpreted such that CNM and diffuse H$_2$ are not so different after all}. This inconsistency may be related to the incomplete mapping of the two clouds with the radio interferometer and perhaps an insufficient angular resolution. With the upcoming HI surveys conducted with Apertif \citep{Oosterloo2010} proper interferometric imaging of large fractions of the sky are expected. These will allow a more detailed and quantitative analysis of the high-Galactic latitude sky and in particular the two IVCs of interest.}

\begin{acknowledgements}
We thank the anonymous referee for many useful comments and suggestions and we acknowledge valuable help of J.~L.~Pineda and P.~Schilke in preparing the observations. We are especially thankful to J.~L.~Pineda who provided us with additional unpublished data. The authors are grateful for the support by the IRAM staff during the observations and for the data reduction. The authors thank the Deutsche Forschungsgemeinschaft (DFG) for financial support under the research grant KE757/11-1. The work is based on observations with the 100$\,$m telescope of the MPIfR (Max-Planck-Institut f{\"u}r Radioastronomie) at Effelsberg. The Planck satellite is operated by the European Space Agency. The development of Planck has been supported by: ESA; CNES and CNRS/INSU-IN2P3-INP (France); ASI, CNR, and INAF (Italy); NASA and DoE (USA); STFC and UKSA (UK); CSIC, MICINN and JA (Spain); Tekes, AoF and CSC (Finland); DLR and MPG (Germany); CSA (Canada); DTU Space (Denmark); SER/SSO (Switzerland); RCN (Norway); SFI (Ireland); FCT/MCTES (Portugal); and PRACE (EU). Some figures have been prepared with the Kapteyn package \citep{KapteynPackage}. T.~R.~is a member of the International Max Planck Research School (IMPRS) for Astronomy and Astrophysics at the Universities of Bonn and Cologne as well as of the Bonn-Cologne Graduate School of Physics and Astronomy (BCGS). 
\end{acknowledgements}

%
%

\bibliographystyle{aa} 
\bibliography{../Literature/Literatur} 

\end{document}